\documentclass[aip,pop,reprint,floatfix]{revtex4-1}

\usepackage{graphicx}
\usepackage{appendix}

\usepackage[utf8]{inputenc}
\usepackage[T1]{fontenc}
\usepackage{mathtools}
\usepackage{cancel}
\usepackage[normalem]{ulem}

\newcommand{\beq}{\begin{equation}}
\newcommand{\eeq}{\end{equation}}

\newcommand{\vu}{\boldsymbol{u}}
\newcommand{\vB}{\boldsymbol{B}}
\newcommand{\vb}{\boldsymbol{b}}

\newcommand{\vA}{v_{\mathrm{A}}}

\newcommand{\dd}{\partial}
\newcommand{\vz}{\boldsymbol{z}}

\newcommand{\ddt}{\dd_t}
\newcommand{\ddx}{\dd_x}

\newcommand{\ddz}{\dd_z}

\newcommand{\vE}{\boldsymbol{E}}
\newcommand{\vJ}{\boldsymbol{J}}

\newcommand{\Dt}{\mathrm{d}_{t}}
\newcommand{\kpar}{k_\parallel}

\newcommand{\rhot}{\rho_\tau}
\newcommand{\nap}{\nabla_\perp}
\newcommand{\kperp}{k_\perp}
\newcommand{\vuperp}{\vu_\perp}
\newcommand{\uz}{u_z}
\newcommand{\vbperp}{\vb_\perp}
\newcommand{\bperp}{b_\perp}
\newcommand{\bz}{b_z}
\newcommand{\hvz}{\hat{\vz}}
\newcommand{\brak}[2]{\left\{#1,#2\right\}}

\newcommand{\vk}{\boldsymbol{k}}
\newcommand{\vrr}{\boldsymbol{r}}

\newcommand{\kpA}{k_{\perp A}}
\newcommand{\kpB}{k_{\perp B}}

\newcommand{\vkpA}{\vk_{\perp A}}
\newcommand{\vkpB}{\vk_{\perp B}}

\newcommand{\kzA}{k_{zA}}
\newcommand{\kzB}{k_{zB}}
\newcommand{\tn}{\tilde{n}}
\newcommand{\kpp}{k_{\perp+}}

\newcommand{\kzp}{k_{z+}}

\draft 
\begin{document}

\title{Nonlinear dynamics of small-scale Alfv\'en waves} 

\author{Alfred Mallet}
\email[Email address for correspondence: ]{alfred.mallet@berkeley.edu}
\affiliation{Space Sciences Laboratory, University of California, Berkeley, CA 94720, USA}
\author{Seth Dorfman}
\affiliation{Space Science Institute, Boulder, CO 80301, USA}
\affiliation{University of California Los Angeles, Los Angeles, California 90095, USA}
\author{Mel Abler}
\affiliation{Space Science Institute, Boulder, CO 80301, USA}
\author{Trevor Bowen}
\affiliation{Space Sciences Laboratory, University of California, Berkeley, CA 94720, USA}
\author{Christopher H.K. Chen}
\affiliation{Department of Physics and Astronomy, Queen Mary University of London, UK}
\date{\today}

\begin{abstract}
We study the nonlinear evolution of very oblique small-scale Alfv\'en waves with $\kperp d_i\gtrsim 1$.
At these scales, the waves become significantly compressive, unlike in MHD, due to the Hall term in the equations.
We demonstrate that when frequencies are small compared to the ion gyrofrequency and amplitudes small compared to unity, no new nonlinear interaction appears due to the Hall term alone at the lowest non-trivial order, even when $\kperp d_i \sim 1$. However, at the second non-trivial order, we discover that the Hall physics leads to a slow but resonant nonlinear interaction between co-propagating Alfv\'en waves, an inherently 3D effect.
Including the effects of finite temperature, finite frequency, and electron inertia, the two-fluid Alfv\'en wave also becomes dispersive once one or more of $\kperp \rho_s$, $\kperp d_e$, or $\kpar d_i$ becomes significant: for oblique waves at low $\beta$ as studied here, this can be at a much smaller scale than $d_i$. We show that the timescale for one-dimensional steepening of two-fluid Alfven waves is only significant at these smaller dispersive scales, and also derive an expression for
the amplitude of driven harmonics of a primary wave. Importantly, both new effects are absent in gyrokinetics and other commonly used reduced two-fluid models.
Our calculations have relevance for the interpretation of laboratory Alfv\'en wave experiments, as well as shedding light on the physics of turbulence in the solar corona and inner solar wind, where the dominant nonlinear interaction between counter-propagating waves is suppressed, allowing these new effects to become important.
\end{abstract}


\maketitle 

\section{Introduction}
Large-amplitude electromagnetic fluctuations are present in a wide range of astrophysical and space plasma physics settings; in the latter case, we have many increasingly precise measurements of these fluctuations from \emph{in situ} spacecraft in the Earth's magnetosphere \citep{torbert2016}, the solar wind \citep{chen2016}, and recently even a spacecraft (Parker Solar Probe, henceforth PSP) in the solar corona \citep{kasper2021}. In the solar wind and especially in the corona, these fluctuations are often polarized (on scales much larger than the ion gyroradius) quite precisely like \emph{Alfv\'en waves} (AW), propagating along the magnetic field in the direction away from the Sun \citep{bale2019,bowen2021}. PSP observations especially have revealed patches of extremely large-amplitude ($\delta B/B\gtrsim 1$) AW, referred to as "switchbacks" since they can even reverse the direction of the background magnetic field. These switchbacks often have remarkably steep boundaries \citep{kasper2019}, suggesting that nonlinear steepening may be an important process in their evolution and/or generation.
\begin{table*}
\centering
\begin{tabular}{|c|c|c|c|c|c|c|c|}
\hline
     Section & $\omega/\Omega_i$ & $\kperp d_i$ & $\kperp \rho_s$ & $\kperp d_e$ & $\beta$ & $\kpar/\kperp$ & $\delta B/B_0$  \\
\hline
     \ref{sec:3DRHMHD} & $\epsilon$ & $1$ & $0$ & $0$ & $0$ & $\epsilon$ & $\epsilon$  \\
\hline
     \ref{sec:1DVDAW}& $1$ & $\epsilon^{-1}$ & $1$ & $1$ & $\epsilon^2$ & $\epsilon$ & $\epsilon$  \\
\hline
     Appendix \ref{app:rdaw}& $\epsilon$ & $\epsilon^{-1}$ & $1$ & $1$ & $\epsilon^2$ & $\epsilon^2$ & $\epsilon^2$  \\

\hline
\end{tabular}
\caption{Summary of the physical regimes studied in different sections of the paper. In each case $\epsilon\ll1$ is a small parameter in which the two-fluid equations are expanded.\label{tab:summary}}
\end{table*}

AW may perform a key role in coronal heating \citep{chandran2009,chandran2021}, due to turbulence driven by reflection from large-scale inhomogeneity in the Alfv\'en velocity $\vA=B_0/\sqrt{4\pi n_{0i} m_i}$. 
In the standard model of MHD turbulence, the nonlinear interaction that permits a turbulent Alfv\'enic cascade is between counterpropagating AW $\vz^\pm = \vu \pm \vb$, where $\vz^\pm$ are the Elsasser variables and $\vu$, $\vb$ are the velocity and magnetic field fluctuations in velocity units\citep{schekochihin2022,howes2015}. However, the PSP observations reveal \citep{bowen2021} that the flux of inward-travelling waves is extremely small compared to that of the dominant outward-travelling AW. The dynamics of this \emph{imbalanced} turbulence may be fundamentally different to balanced turbulence with a comparable flux of AW in both directions, allowing different physics to dominate the dynamics of the system. For example, in an important theoretical development, it has recently been shown that the conservation of generalized helicity causes a barrier to turbulent energy flux to appear at the ion scales in imbalanced turbulence \citep{meyrand2021}, leading to remarkably large-amplitude fluctuations and enabling high-frequency ion-cyclotron heating of the plasma \citep{squire2021}, which allows for the necessary perpendicular ion heating to power the fast solar wind. This provides some motivation for revisiting the physics of small-scale AW, including nonlinear steepening and co-propagating nonlinear interactions.

Alfv\'en waves are also routinely launched in laboratory experiments at the LArge Plasma device (LAPD), which is a cylindrical device that produces a $16.5\mathrm{m}$-long column of quiescent, magnetized plasma \citep{LAPD}. Waves with various parameters can be launched using antennae placed at the ends of the column; there have been extensive experiments performed using LAPD on the linear \citep{leneman1999,gekelman2000,vincena2004,palmer2005} and nonlinear \citep{carter2006,auerbach2010,howes2012,dorfman2013,dorfman2015,dorfman2016} properties of AW. The plasma regime in which these waves are launched may be similar to small-scale waves present in the solar corona and Earth's aurora: $\kperp d_i \gg1$ while $\kpar d_i\lesssim 1$, but $\kperp\rhot\sim \kperp d_e \lesssim 1$, implying $\kpar/\kperp\ll1$ and $\beta \sim Zm_e/m_i\ll1$, where $\kpar$ ($\kperp$) are the components of the wavenumber $k$ parallel (perpendicular) to the mean magnetic field , $d_i=\vA/\Omega_i$ the ion inertial length, $\rhot=c_s/\Omega_i$ the ion sound radius, $d_e=\sqrt{Zm_e/m_i}d_i$ the electron inertial length, and $\beta=c_s^2/\vA^2$, with the Alfv\'en velocity $\vA=B_0/\sqrt{4\pi n_{0i}m_i}$, sound velocity $c_s=\sqrt{(T_i+ZT_e)/m_i}$, and ion gyrofrequency $\Omega_i=ZeB_0/m_ic$. Unlike the corona, the electron-ion collision frequency in the device can be quite large, $\nu_{ei}\sim\omega$ where $\omega\sim\kpar\vA$ is the frequency of the waves; thus, for this setting it is important to retain the Ohmic resistivity $\eta = \nu_{ei}d_e^2$. Interestingly, harmonics are routinely observed in AW experiments on LAPD\citep{brugman2007,dorfman2016,abler2023}, even without a counterpropagating wave, suggesting that improved modelling of AW nonlinear behaviour is needed to understand the experimental results, as well as the aforementioned coronal turbulence.

Within the framework of compressible magnetohydrodynamics (MHD), the Alfv\'en wave is an exact nonlinear solution, for which velocity and magnetic field fluctuations obey $\delta \vu = \pm \delta \vB/\sqrt{4\pi\rho}$, and additionally the magnetic-field strength, density and pressure are space-time constants. This situation is known as spherical polarization since the magnetic-field vector moves on the surface of a sphere: it should be noted that this means that the large-amplitude AW is explicitly not monochromatic \citep{goldstein1974,barnes1974}. Such a configuration propagates at $\vA$ without steepening of the perpendicular fluctuations, regardless of amplitude or three-dimensional structure\citep{goldstein1974}. There are parallel fluctuations $\delta B_\parallel = -\delta B_\perp^2/2B_0$, required to enforce the space-time constancy of $|\vB|$, appearing as harmonics of the perpendicular fluctuations\citep{barnes1974,vasquez1998a}: however, they do not depend on time in this solution and their amplitude is much smaller than the harmonic amplitudes we will derive in this paper. For MHD Alfv\'en waves, non-trivial nonlinear interaction only occurs due to the presence of counter-propagating waves.

Even at large amplitude, the MHD Alfv\'en wave is precisely incompressible. Going beyond MHD, including the Hall effect in Ohm's law (see Appendix \ref{app:2f} for a brief derivation of relevant equations), very oblique ($\kperp \gg \kpar$) AW develop density fluctuations\citep{hollweg1999},
\beq
\frac{\delta n_i}{n_{0i}} \sim - i \kperp d_i \frac{\delta B}{B_0},\label{eq:densityintro}
\eeq
scaling with $\kperp d_i$, along with a corresponding compressive flow (the ion polarization drift). The wave only becomes \emph{dispersive} when one of $\kperp \rhot, \kperp d_e, \kpar d_i$ becomes significant: the linear two-fluid dispersion relation in an appropriate limit, including the effects of resistive damping, is (see Appendix~\ref{app:2f} and Hollweg 1999\cite{hollweg1999})
\beq
(1 + k_\parallel^2d_i^2 + k^2 d_e^2)\omega^2 + i\eta k^2\omega - k_\parallel^2\vA^2(1+k^2\rhot^2) = 0,\label{eq:disp}
\eeq
with solution
\beq
\omega = -i\gamma_0 \pm \sqrt{\omega_0^2-\gamma_0^2},
\eeq
where
\beq
\omega_0 = k_\parallel \vA \sqrt{\frac{1+k^2\rhot^2}{1+k_\parallel^2d_i^2+k^2d_e^2}},\quad \gamma_0 = \frac{1}{2}\frac{\eta k^2}{1+k_\parallel^2 d_i^2 + k^2 d_e^2}.
\eeq
While not always accurate (e.g. when kinetic effects become important, see Appendix \ref{app:2f} for more discussion), the two-fluid system thus provides a minimal model for studying dispersive AW. For oblique waves at low $\beta$, the dispersive terms only become relevant at much smaller scales than $d_i$\citep{hollweg1999}, so that if (for example) $\kperp\rhot$ is the largest of these dispersive parameters, between $\rhot\ll 1/\kperp \lesssim d_i$ the wave is compressive but not significantly dispersive 
(this conclusion holds for a system that is not too collisional, $\nu_{ei}\lesssim\omega$, as the resistive damping then only enters when $\kperp d_e \sim 1$). Due to the density fluctuations, once $\kperp d_i \sim 1$ the local Alfven velocity $v_A'=B/\sqrt{4\pi n_{i} m_i}$ now varies with the wave's phase: at sufficiently large amplitude, it might be natural to assume that the Hall term might drive new nonlinear interactions and steepening, even between co-propagating waves. These new interactions are the subject of this paper.

First, in Section ~\ref{sec:3DRHMHD}, we derive a set of equations describing three-dimensional dynamics of very oblique AW for $\kperp d_i \sim 1$, but assuming that $\rho_s = d_e =0$ and $\omega/\Omega_i \ll 1$, so that the linear dispersion relation (in this oblique limit) is just $\omega = \pm \kpar \vA$. This results in an reduced expansion of the Hall MHD equations. At the lowest non-trivial order, we reproduce the classic equations of reduced MHD (RMHD), in which nonlinearity only occurs between counter-propagating AW, despite the significant density fluctuations: a previously known result (see, e.g. Zocco \& Schekochihin 2011\citep{zocco2011}, or Appendix \ref{app:rdaw}): thus, at this order, the Hall effect makes no difference to the dynamics. However, at the next order in our asymptotic expansion, a three-dimensional nonlinear interaction does exist between co-propagating waves, despite no change in the linear dispersion relation at this order either. Because this effect only appears at next order, the timescale associated with the interaction is slow compared to the Alfv\'en timescale $\tau_A = (\kpar\vA)^{-1}$. Moreover, this interaction vanishes if the perpendicular wavevectors of two interacting waves are aligned, e.g. in a one-dimensional system.

Once the AW becomes dispersive (e.g. when $\kperp \rhot$ or $\kperp d_e$ can no longer be neglected), it is already known nonlinear interactions between co-propagating waves are possible, even at small amplitude and low frequency \citep{seyler2003,schektome2009,zocco2011}: waves of different $\kperp$ "catch up" with each other and interact. This well-known nonlinearity is not the main focus of this paper, but we provide a systematic derivation of relevant equations in Appendix~\ref{app:rdaw}. Again, this nonlinear interaction vanishes if the perpendicular wavevectors of the parent waves are aligned.

Both of the aforementioned nonlinear interactions vanish in one dimension. We show in Sec.~\ref{sec:1DVDAW} that one-dimensional steepening of two-fluid AW is rather unusual: it requires both large density fluctuations and significant dispersive terms to be active. If all the dispersive terms and resistivity are negligible, even with large density fluctuations ($\kperp d_i\gg1$), in one dimension the compressive flow exactly cancels the modification to the local Alfv\'en velocity, resulting in no net steepening. We derive an expression for the harmonic of the primary AW driven by dispersive steepening, which can be compared to harmonics observed in laboratory experiments on LAPD \citep{abler2023}. 

In a situation with equal fluxes of counterpropagating AW, the new nonlinear interactions discovered in Secs.~\ref{sec:3DRHMHD} and \ref{sec:1DVDAW} would be subdominant for $\kperp d_i\sim 1$. The timescales associated with both processes scale with the amplitude of the density fluctuations (Eq.~\ref{eq:densityintro}). However, the counterpropagating interaction is suppressed in the highly imbalanced, regime in the solar corona, where moreover $\beta\ll1$: this may allow these effects to become important in the transition between the inertial-range and kinetic-range turbulence, via a competition between the level of imbalance $R_E= \delta z^-/\delta z^+$ and the smallness of the density fluctuations. Moreover, the existence of these interactions means that care must be taken interpreting laboratory experiments \citep{howes2012,abler2023} studying nonlinear AW interactions on LAPD, where $\kperp d_i \gg1$, shortening the nonlinear timescales associated with these new processes.

The different physical regimes studied in the paper are summarized in Table \ref{tab:summary}. The three ordering schemes here are designed to isolate the three different types of co-propagating nonlinear interaction: three-dimensional Hall-driven interaction (Sec. \ref{sec:3DRHMHD}), one-dimensional steepening, which requires both density fluctuations and the waves to be dispersive (Sec. \ref{sec:1DVDAW}, and purely dispersively-driven nonlinearity (Appendix \ref{app:rdaw}, already known e.g. from previous reduced fluid\cite{seyler2003} and/or gyrokinetic models\cite{schektome2009,zocco2011}). Taken together, our results shed new light on the physics of Alfv\'en waves in laboratory and space plasmas, and may have relevance to both existing experiments and on the evolution of extremely imbalanced turbulence and switchbacks in the corona and solar wind.

\section{Reduced Hall MHD to next order: 3D nonlinear interaction}\label{sec:3DRHMHD}
As pointed out in the introduction, for $\beta\ll1$, $\nu_{ei} \lesssim\omega$, and $ \kpar/\kperp \ll 1$, AW become compressive due to the Hall term at $\kperp d_i \sim 1$, a much larger scale than those at which the linear dispersion relation (\ref{eq:disp}) differs from ideal MHD. Are the nonlinear properties similarly unaffected? This question is the subject of this section. 

The essential new physics will appear even though from the outset we completely neglect the kinetic and inertial AW dispersive terms and the resistivity (in Eqs.~\ref{eq:cty2}--\ref{eq:gauss2}, take $c_s/v_A\to 0$, $m_e/m_i\to 0$, and $\eta \to 0$), leaving just the equations of compressible Hall MHD:
\begin{align}
    \ddt n &=- \nabla\cdot(n\vu),\label{eq:hcty}\\
    \Dt \vu &= \frac{1}{n}(\nabla \times \vb)\times\vb,\label{eq:hmom}\\
    \ddt \vb &= \nabla\times(\vu \times\vb) - d_i\nabla\times\Dt\vu,\label{eq:hind}\\
    \nabla\cdot\vb &= 0,\label{eq:hgauss2}
\end{align}
where $\vu$ is the ion velocity, $n=n_i/n_{0i}$, and the magnetic field in Alfv\'en units is $\vb = \vB/\sqrt{4\pi n_{0i}m_i}$. We write
\beq
\vb = \vbperp + \hvz\bz + \hvz\vA, \quad n = 1 + \tn, \quad \vu = \vuperp + \hvz \uz.
\eeq
where the Alfv\'en speed $\vA=B_0/\sqrt{4\pi n_{0i}m_i}$ is the magnitude of the (straight) mean magnetic field, taken to be in the $\hvz$ direction, and $\vb_\perp$, $\bz$, $\tn$, $\vuperp$, and $\uz$ are fluctuations. 
We assume that the waves are highly oblique, and use as our expansion parameter
\beq
\epsilon \sim \frac{\ddz}{\nap} \ll1.
\eeq
We also assume that the linear timescales of our fluctuations are Alfv\'enic, and comparable to the nonlinear timescales,
\beq
\ddt \sim \vA\ddz \sim \vuperp\cdot \nap \sim \vbperp \cdot \nap,
\eeq
whence
\beq
\frac{u_\perp}{\vA}\sim \frac{\bperp}{\vA} \sim \epsilon.
\eeq
This means that the critical balance parameter\citep{gs95}
\beq
\chi= \frac{\kperp\bperp}{\kpar\vA} \sim 1\label{eq:chi}
\eeq
is permitted to be of order unity. Finally, we also take the perpendicular scale of the fluctuations to be comparable to $d_i$
\beq
d_i \nap\sim 1,
\eeq
implying that the frequencies are small compared to $\Omega_i$,
\beq
d_i\ddz \sim \frac{1}{\Omega_i}\ddt \sim \epsilon.
\eeq
Some of these orderings may be somewhat familiar from reduced MHD \citep{strauss1976,schektome2009}.
For the other fluctuations, we write
\beq
\tn \sim \epsilon^{\alpha_n}, \quad \frac{\nap \cdot \vuperp}{\Omega_i}\sim \epsilon^{\alpha_\xi}, \quad \frac{b_z}{\vA} \sim \epsilon^{\alpha_b}, \quad \frac{\uz}{\vA} \sim \epsilon^{\alpha_u},
\eeq
where we will deduce the values of $\alpha_n, \alpha_\xi, \alpha_b, \alpha_u$ in the following analysis. We will keep terms at the lowest \emph{two} non-trivial orders in each equation. Note that the dispersive correction to the linear AW frequency from finite $\omega/\Omega_i$ (equivalently finite $\kpar d_i$) is $\sim (\omega/\Omega_i)^2\kpar \vA\sim \epsilon^3\Omega_i$ (see Eq.\ref{eq:disp}), and so will be neglected, meaning that the linear dispersion relation in our system up to the lowest two orders is just $\omega = \pm \kpar \vA$.

We now order the terms in our equations; the orderings shown below the equations are obtained by normalizing times by $\Omega_i^{-1}$, velocities and magnetic field in Alfv\'en units by $\vA$, and lengths by $d_i = \vA/\Omega_i$.

The perpendicular components of (\ref{eq:hmom}) are
\begin{align}
\underbrace{\ddt \vuperp}_{\epsilon^2} &+ \underbrace{\vuperp\cdot\nap \vuperp}_{\epsilon^2} + \cancel{\underbrace{\uz\ddz \vuperp}_{\epsilon^{2+\alpha_u}}} \nonumber\\= \frac{1}{n}&\left[\underbrace{\vA\ddz\vbperp}_{\epsilon^2} + \underbrace{\vbperp\cdot\nap\vbperp}_{\epsilon^2} + \cancel{\underbrace{\bz\ddz\vbperp}_{\epsilon^{2+\alpha_b}}}\right.\nonumber\\& \left.- \nap\left(\underbrace{\frac{1}{2}\bperp^2}_{\epsilon^2} + \underbrace{\bz\vA}_{\epsilon^{\alpha_b}} +\cancel{\underbrace{\bz^2}_{\epsilon^{2\alpha_b}}}\right)\right],\label{eq:perpmom}
\end{align}
where, according to the normalizations in the previous paragraph, the orderings shown below each term are relative to $\Omega_i\vA$. For the $\nap b_z\vA$ term to not be larger than all others, $\alpha_b \geq 2$, so we take
\beq
\frac{b_z}{\vA} \sim \epsilon^2,
\eeq
which allows us to neglect some of the pre-emptively crossed-out terms in (\ref{eq:perpmom}). The $\hvz$-component of (\ref{eq:hmom}) is 
\begin{align}
\underbrace{\ddt \uz}_{\epsilon^{1+\alpha_u}} &+ \underbrace{\vuperp\cdot\nap\uz}_{\epsilon^{1+\alpha_u}} + \cancel{\underbrace{\uz\ddz\uz}_{\epsilon^{1+2\alpha_u}}} \nonumber\\&= \frac{1}{n}\left[\underbrace{\vbperp \cdot \nap \bz}_{\epsilon^{3}} + \cancel{\underbrace{\bz\ddz\bz}_{\epsilon^{5}}} - \underbrace{\ddz\left(\frac{1}{2}\bperp^2\right)}_{\epsilon^{3}}\right].\label{eq:zmom}
\end{align}
This implies $\alpha_u \geq 2$, and so we take
\beq
\frac{u_z}{\vA} \sim \epsilon^2,
\eeq
allowing us to neglect some more of the crossed-out terms in (\ref{eq:perpmom}) and (\ref{eq:zmom}).
The $\hvz$-component of (\ref{eq:hind}) gives
\begin{align}
\underbrace{\nap\cdot \vuperp}_{\epsilon^{\alpha_\xi}} = &-\underbrace{\frac{1}{\Omega_i}\nap\times\Dt\vuperp}_{\epsilon^2} \nonumber\\&+ \underbrace{\frac{1}{\vA}\nap\cdot(\uz\vbperp - \bz\vuperp)}_{\epsilon^3}-\frac{1}{\vA}\underbrace{\ddt\bz}_{\epsilon^3},\label{eq:hpoldrift}
\end{align}
We can deduce that $\alpha_\xi\geq 2$, and so we take
\beq
\frac{\nap\cdot\vuperp}{\Omega_i} \sim \epsilon^2.
\eeq
The different terms in (\ref{eq:hcty}) are then ordered as
\beq
\underbrace{\ddt \tn}_{\epsilon^{1+\alpha_n}} + \underbrace{\vuperp\cdot\nap\tn}_{_{\epsilon^{1+\alpha_n}}} + \cancel{\underbrace{\uz\ddz\tn}_{\epsilon^{3+\alpha_n}}} + n(\underbrace{\nap\cdot\vuperp}_{\epsilon^2} + 
\underbrace{\ddz\uz}_{\epsilon^3}) = 0,\label{eq:ctyordered}
\eeq
so that $\alpha_n\geq 1$ and we take
\beq
\tn \sim \epsilon.
\eeq
We may then drop the crossed-out term in (\ref{eq:ctyordered}) since it is two orders below the leading terms.

Now, note that (\ref{eq:hgauss2}) gives
\beq
\underbrace{\nap\cdot \vbperp}_{\epsilon} + \underbrace{\ddz\bz}_{\epsilon^3} = 0.
\eeq
Since the second term is two orders down from the first, we may write
\beq
\vbperp = \hvz \times\nap \psi + O(\epsilon^3\vA),
\eeq
as in (\ref{eq:psidef}), so that at the lowest two orders in our expansion
\beq
\nap\cdot\vbperp = 0, \quad \nap\times\vbperp = \nap^2\psi.
\eeq
We will also find it convenient to decompose the two independent components of $\vuperp$ as
\beq
\vuperp = \hvz\times\nap \phi + \nap\xi,
\eeq
so that
\beq
\nap\cdot\vuperp = \nap^2\xi, \quad \nap\times\vuperp = \nap^2\phi.
\eeq
In terms of these scalar variables, up to the second non-trivial order we have
\begin{align}
\Dt f &= \underbrace{\ddt f + \{\phi,f\}}_{\epsilon} + \underbrace{\nap\xi\cdot \nap f}_{\epsilon^2}, \\
\vb \cdot \nabla f &= \underbrace{\vA\ddz f + \{\psi,f\}}_{\epsilon},
\end{align}
where $\{f,g\} = \nap f \times \nap g$, $f$ and $g$ being arbitrary functions, and the orderings are relative to $\Omega_i f$.

Operating on (\ref{eq:perpmom}) with $\nap\times$, we find
\begin{align}
&\underbrace{\ddt \nap^2\phi + \{\phi,\nap^2\phi\}}_{\epsilon^2} + \underbrace{\nap\cdot(\nap^2\phi\nap\xi)}_{\epsilon^3} \nonumber\\&= \underbrace{\vA\nap\cdot\left(\frac{1}{n}\ddz\nap\psi\right) + \left\{\psi,\frac{1}{n}\nap^2\psi\right\}}_{\epsilon^2}+\underbrace{\left\{\vA\bz,\frac{1}{n}\right\}}_{\epsilon^3}.\label{eq:curlmom}
\end{align}
Operating on (\ref{eq:perpmom}) instead with $\nap\cdot$ yields
\begin{align}
&\underbrace{\ddt \nap^2\xi}_{\epsilon^3} + \frac{1}{2}\nap^2\left(\underbrace{|\nap\phi|^2}_{\epsilon^2}+\cancel{\underbrace{|\nap\xi|^2}_{\epsilon^4}}\right) \nonumber\\&- \underbrace{\nap\cdot\left(\nap^2\phi\nap\phi\right)}_{\epsilon^2}+\underbrace{\nap^2\{\phi,\xi\}+\{\xi,\nap^2\phi\}}_{\epsilon^3}\nonumber\\
&=\underbrace{\vA\left\{\ddz\psi,\frac{1}{n}\right\}}_{\epsilon^3}-\underbrace{\nap\cdot\left(\frac{1}{n}\nap^2\psi\nap\psi + \frac{1}{n}\nap\bz\vA\right)}_{\epsilon^2}.\label{eq:divmom}
\end{align}
Operating on the perpendicular components of (\ref{eq:hind}) with $\nap\times$, we obtain
\beq
\nap^2\Dt \psi = \vA\ddz\nap^2\phi + d_i[\nap^2\ddt \uz - \ddz\nap\cdot\Dt\vuperp].
\eeq
Using the lowest orders of (\ref{eq:divmom}) and (\ref{eq:zmom}), we can write this as
\begin{align}
\underbrace{\nap^2\Dt\psi}_{\epsilon^2} = &\underbrace{\vA\ddz\nap^2\phi}_{\epsilon^2} \nonumber\\&+ \underbrace{d_i \nap^2\left(\{\psi,\bz\} - \frac{1}{2}\ddz\left(|\nap\psi|^2\right)\right)}_{\epsilon^3}\nonumber\\&+\underbrace{d_i \ddz\nap\cdot\left(\nap^2\psi\nap\psi+\nap\bz\vA\right)}_{\epsilon^3},\label{eq:curlind}
\end{align}
where we have neglected all higher-order terms. (\ref{eq:ctyordered}), (\ref{eq:hpoldrift}), (\ref{eq:zmom}), (\ref{eq:divmom}), (\ref{eq:curlmom}), and (\ref{eq:curlind}) constitute a closed set of equations for our six variables $\phi,\psi,\xi,\bz,\uz,\tn$. We will now proceed to expand systematically in $\epsilon$, writing $\phi = \epsilon\phi_1 + \epsilon^2\phi_2$, and so on for the other variables, taking into account their overall ordering, e.g. $\bz = \epsilon^2 b_{z1} + \epsilon^3 b_{z2}$.
\subsection{Lowest order: RMHD$+$}
If we keep only the lowest-order terms, Eqs.~(\ref{eq:curlmom}) and (\ref{eq:curlind}) reduce to the equations of RMHD,
\begin{align}
\ddt \nap^2\phi_1 + \{\phi_1,\nap^2\phi_1\} &= \vA\ddz\nap^2\psi_1 + \left\{\psi_1,\nap^2\psi_1\right\},\label{eq:phi1}\\
\ddt\psi_1 + \{\phi_1,\psi_1\} &= \vA\ddz\phi_1,\label{eq:psi1}
\end{align}
i.e. the dynamics of the Alfv\'en wave is at this order unaffected by the fact that $\kperp d_i \sim 1$. The lowest orders of the other equations yield
\begin{align}
\tn_1 &= \frac{1}{\Omega_i}\nap^2\phi_1,\label{eq:tn1}\\
\nap^2\xi_1 &= -\frac{1}{\Omega_i}\Dt\nap^2\phi_1,\label{eq:xi1}\\
\nap^2b_{z1}\vA &= \nap\cdot\left(\nap^2\phi_1\nap\phi_1-\nap^2\psi_1\nap\psi_1\right)\nonumber\\&\quad-\frac{1}{2}\nap^2|\nap\phi_1|^2,\label{eq:bz1}\\
\Dt u_{z1} &= \{\psi_1, b_{z1}\}-\frac{1}{2}\ddz|\nap\psi_1|^2,\label{eq:uz1}
\end{align}
It is worth noticing that
\beq
\phi_1 = \phi_1(x,y,z\pm\vA t) = \pm \psi_1
\eeq
constitutes an exact nonlinear solution at this order, an Alfv\'en wave with arbitrary perpendicular structure propagating antiparallel ($+$) or parallel ($-$) to the background magnetic field at group velocity $\vA$. For this solution, we obtain from (\ref{eq:bz1}) and (\ref{eq:uz1})
\beq
b_{z1} = - \frac{\bperp^2}{2\vA} = \pm u_{z1},
\eeq
and the total magnetic-field-strength-squared is constant to lowest order, a generalization of the MHD large-amplitude AW\citep{barnes1974,vasquez1998a,vasquez1998b}.
\subsection{Next order}
To obtain the equations to the next order, we must remember to take into account the fact that $n = 1+ \epsilon \tn_1 + O(\epsilon^2)$, the second-order pieces of the time derivative $\Dt f$,
\beq
\left[\ddt f_2  + \{\phi_2,f_1\} + \{\phi_1,f_2\} + \nap\xi_1\cdot\nap f_1\right],
\eeq
and all the next-order terms that are already written explicitly in the preceding equations. Solving for $n_2, \xi_2,b_{z2}$, and $u_{z2}$ will not be necessary. The final two equations, (\ref{eq:curlmom}) and (\ref{eq:curlind}), are at this order
\begin{align}
&\ddt\nap^2\phi_2 -\vA\ddz\nap^2\psi_2 \nonumber\\&+ \{\phi_2,\nap^2\phi_1\} + \{\phi_1,\nap^2\phi_2\}-\{\psi_2,\nap^2\psi_1\} - \{\psi_1,\nap^2\psi_2\}\nonumber\\
&=\underbrace{-\nap\cdot\left[\nap^2\phi_1\nap\xi_1+ \tn_1\vA\ddz\nap\psi_1\right]}_{A}\nonumber\\
&\quad\underbrace{-\{\psi_1,\tn_1\nap^2\psi_1\}+\vA\{\tn_1,b_{z1}\}}_{B},\label{eq:curlmomnext}\\
&\ddt\nap^2\psi_2 -\vA\ddz\nap^2\phi_2+ \nap^2(\{\phi_1,\psi_2\}+\{\phi_2,\psi_1\})\nonumber\\
&= \underbrace{d_i\nap^2\brak{\psi_1}{b_{z1}}}_{C}\underbrace{-\nap^2\left[\frac{1}{2}d_i\ddz|\nap\psi_1|^2+\nap\xi_1\cdot\nap\psi_1\right]}_{D}\nonumber\\&\quad+\underbrace{d_i\left[\ddz\nap\cdot\left(\nap^2\psi_1\nap\psi_1+\nap b_{z1}\vA\right)\right]}_{E}.\label{eq:curlindnext}
\end{align}
The evolution of $\phi_2$ and $\psi_2$ depend only on themselves and the lowest-order variables. The labelled terms that comprise the RHS of each equation only depend on the lowest-order fluctuations, i.e. they drive $\phi_2$ and $\psi_2$. Using (\ref{eq:xi1}) combined with (\ref{eq:phi1}) to substitute for $\xi_1$, and (\ref{eq:tn1}) to substitute for $\tn_1$,
\beq
A = \frac{1}{\Omega_i}\nap\cdot\left[\nap^2\phi\nap \nap^{-2}\{\psi_1,\nap^2\psi_1\}\right],
\eeq
where $\nap^{-2}$ is the inverse operator to $\nap^2$, and we cancel some terms. Next,
\begin{align}
B = &-\frac{1}{\Omega_i}\left[\nap^2\phi_1\brak{\psi_1}{\nap^2\psi_1}+\nap^2\psi_1\brak{\psi_1}{\nap^2\phi_1}\right.\nonumber\\&\phantom{\frac{1}{\Omega_i}}\left.-\brak{\nap^2\phi_1}{b_{z1}\vA}\right],
\end{align}
$C$ does not benefit from any manipulation. Using (\ref{eq:xi1}) combined with (\ref{eq:phi1}) to substitute for $\xi_1$, and cancelling two terms, we find that
\beq
D=\frac{1}{\Omega_i}\nap^2\left[\nap\psi_1\cdot\nap\nap^{-2}\brak{\psi_1}{\nap^2\psi_1}\right]
\eeq
Finally, using (\ref{eq:bz1}), we find
\begin{align}
E &= d_i\ddz \nap\cdot\left[\nap^2\phi_1\nap\phi_1-\frac{1}{2}\nap|\nap\phi_1|^2\right]\nonumber\\
&=d_i\ddz \nap\cdot\left[(\hvz\times\nap\phi_1)\cdot\nap(\hvz\times\nap\phi_1)\right].
\end{align}
Notably, all of $A$-$E$ vanish if all perpendicular gradients in the system are aligned. $A$-$D$ involve three powers of the lowest-order AW amplitude (remembering that $b_{z1}$ is already nonlinear, given by Eq.~\ref{eq:bz1}), while $E$ involves only two; thus, $E$ would dominate if we performed a subsidiary expansion in $\chi\ll1$ (Eq.~\ref{eq:chi}); for $\chi\sim1$, all terms are comparable.
Defining \citealt{elsasser1950} variables
\beq
\zeta^\pm = \phi\pm\psi,\label{eq:elsasser}
\eeq
and taking the sum and difference of (\ref{eq:curlmomnext}) and (\ref{eq:curlindnext}), our equations may be written
\begin{widetext}
\begin{align}
&\ddt\nap^2\zeta_2^\pm\mp\vA\ddz\nap^2\zeta_2^\pm \nonumber\\
&=-\frac{1}{2}\left[\brak{\zeta_2^-}{\nap^2\zeta_1^+}+\brak{\zeta_2^+}{\nap^2\zeta_1^-}+\brak{\zeta_1^-}{\nap^2\zeta_2^+}+\brak{\zeta_1^+}{\nap^2\zeta_2^-}\mp\nap^2\left(\brak{\zeta_1^+}{\zeta_2^-}+\brak{\zeta_2^+}{\zeta_1^-}\right)\right]\nonumber\\
&\quad
+\frac{1}{\Omega_i}\left[\nap\nap^2\phi_1\cdot \nap \nap^{-2}\brak{\psi_1}{\nap^2\psi_1}
-\nap^2\psi_1\brak{\psi_1}{\nap^2\phi_1}+\brak{\nap^2\phi_1}{b_{z1}\vA}\right]\nonumber\\
&\quad\pm\frac{1}{\Omega_i}\left[
\nap^2\brak{\psi_1}{b_{z1}\vA}
-\nap^2\left(\nap\psi_1\cdot\nap\nap^{-2}\brak{\psi_1}{\nap^2\psi_1}\right)
+\vA\ddz\nap\cdot\left((\hvz\times\nap\phi_1) \cdot \nap (\hvz\times\nap\phi_1)\right)
\right]\label{eq:els2ord}
\end{align}
\end{widetext}
The first line of the RHS are just the standard RMHD nonlinearities between $\zeta_2^\pm$ and $\zeta_1^\mp$. The second and third lines of the RHS describe a nonlinear drive for $\zeta_2^\pm$. Most of these terms are cubic in the fluctuation amplitude, but the final term is quadratic, and thus dominates when $\chi\ll1$.

The result of this calculation may be summarized as follows: at lowest nontrivial order in the RMHD expansion arbitrary 3D structure is permitted for Alfv\'en waves travelling in only one direction, even when $d_i\nap\sim 1$. However, at next order in the expansion, new terms appear which break this nonlinear solution: taking $\zeta_1^-=0$, these new nonlinear terms drive both $\zeta_2^+$ and $\zeta_2^-$. 
\subsection{Co-propagating three-wave interaction, $\chi\ll1$}\label{sec:3wave}
Let us suppose that
\begin{align}
&\zeta_1^-=0, \quad \zeta_1^+ = \zeta_{A}^+ + \zeta_{B}^+, \nonumber\\ &\zeta^+_{A,B} = s_{A,B} e^{i(\vk_{A,B}\cdot \vrr +k_{zA,B}\vA t)}+ \text{c.c.},
\end{align}
which is an exact nonlinear solution at lowest order (i.e. in RMHD), with 
\beq
|s_{A,B}| = \frac{\chi_{A,B}\vA k_z}{\kperp^2},
\eeq
and $\chi_{A,B}\ll1$. At second order in this weakly nonlinear case, all but the last term on the RHS of (\ref{eq:els2ord}) may be neglected since they are smaller by a factor $\chi$, and since we have taken $\zeta^-_1=0$ we are left with
\begin{align}
&\ddt\nap^2\zeta_2^\pm\mp\vA\ddz\nap^2\zeta_2^\pm\nonumber\\&=\pm\frac{1}{4}d_i\ddz\nap\cdot\left((\hvz\times\nap\zeta_1^+) \cdot \nap (\hvz\times\nap\zeta_1^+)\right).
\end{align}
Note that if $\zeta^-_1\neq0$, the term will also have contributions from counter-propagating waves; however, in the case where there are counterpropagating waves this interaction would be subdominant to the RMHD nonlinearity at lower order. 
The RHS has nonzero components with two different phase functions
\beq
(\vk_A\pm\vk_B)\cdot \vrr + (k_{zA}\pm k_{zB})\vA t,
\eeq
and we may write the resonant solution $\zeta_2^\pm = \zeta_{A+B}^\pm + \zeta_{A-B}^\pm$ where
\begin{align}
\zeta_{A+B}^\pm &= s_{A+B}^\pm e^{i((\vk_A+\vk_B)\cdot\vrr + (k_{zA}+k_{zB})\vA t)}+ \text{c.c.},\\
\zeta_{A-B}^\pm &= s_{A-B}^\pm e^{i((\vk_A-\vk_B)\cdot\vrr + (k_{zA}-k_{zB})\vA t)}+\text{c.c.}
\end{align}
It is not necessary to solve for both of these, so we choose $A+B$. Substituting in for $\zeta_1^+$ on the RHS and matching the spatial dependence of both sides,
\begin{align}
&-\kpp^2\left[\ddt \zeta_{A+B}^\pm \mp i\kzp\vA\zeta_{A+B}^\pm\right]\nonumber\\&=\pm\frac{1}{2}i\kzp d_i |\vkpA\times\vkpB|^2 s_A s_B e^{i(\vk_+\cdot r + \kzp \vA t)} + \text{c.c.},
\end{align}
where $\vk_+= \vk_A + \vk_B$. The solution is a non-resonant $\zeta^-$ mode with amplitude
\beq
s_{A+B}^- = \frac{s_As_B}{4\Omega_i}\frac{|\vkpA\times\vkpB|^2}{|\vkpA+\vkpB|^2},
\eeq
and a resonant $\zeta^+$ mode,
\beq
s^+_{A+B} = - \left[\frac{1}{2}i (k_{zA}+k_{zB}) d_i \frac{|\vkpA\times\vkpB|^2}{|\vkpA+\vkpB|^2} s_A s_B\right]t,
\eeq
with the solution only valid at sufficiently early times that $s_{A+B}^+/s_{A,B} \ll 1$. Writing these in terms of the magnitudes of vector Elsasser variables $\vz^\pm = \vz \times \nap \zeta^\pm$,
\begin{align}
    \frac{z^-_{A+B}}{\vA} &= \frac{1}{4} \left(\frac{\kpA\kpB}{|\vkpA+\vkpB|}\right)d_i \sin^2\alpha \frac{z^+_A z^+_B}{\vA^2}, \label{eq:zminusamp}\\
    \frac{z^+_{A+B}}{\vA} &= -\frac{1}{2}\left(\frac{\kpA\kpB}{|\vkpA+\vkpB|}\right)d_i\sin^2\alpha\frac{z^+_A z^+_B}{\vA^2} (\kzA+\kzB)\vA t,
\end{align}
where $\alpha$ is the angle between $\kpA$ and $\kpB$. Supposing that the two primary waves have roughly similar scales and amplitudes, i.e. $\kpA \sim \kpB\sim \kpp$, $\kzA\sim\kzB\sim\kzp$, and $z^+_A\sim z^+_B \sim z^+_1$, in terms of scalings
\begin{align}
\frac{z^-_{2}}{\vA}&\sim \kperp d_i \sin^2\alpha \left(\frac{z^+_1}{\vA}\right)^2, \\
\frac{z^+_{2}}{\vA} &\sim \kperp d_i \sin^2\alpha \left(\frac{z^+_1}{\vA}\right)^2 k_z \vA t.
\end{align}
The calculation breaks down once $z^+_2 \sim z^+_1$, on a timescale
\beq
\tau_{H2} \sim \tau_A(\kperp d_i \sin^2\alpha z^+_1/\vA)^{-1},\label{eq:tauh2}
\eeq
where $\tau_A = 1/k_z\vA$; as expected, $\tau_{H2}\gg \tau_A$ by a factor of $1/\epsilon$. More specifically, it is slower by a factor $\sim n_1$, the density fluctuations (see Eq.~\ref{eq:densityintro}): we will study a case where the density fluctuations are permitted to become large in Sec.~\ref{sec:1DVDAW}.

\subsection{Strong nonlinearity, $\chi \gtrsim 1$}\label{sec:strong}
When $\chi\gtrsim 1$ (as is likely the case in turbulence in the corona), the cubic nonlinear terms in (\ref{eq:els2ord}) are significant; moreover, the driven wave is large enough that the RMHD-like terms on the first line of the RHS of (\ref{eq:els2ord}) are also important: resonant $n$-wave interactions are comparable for all $n\geq3$. Another nonlinear timescale may be defined by comparing the cubic nonlinear terms with the time-derivative term,
\beq
\tau_{H3} \sim \left(\kperp d_i |\sin\alpha|\kperp (z^+)^2/{\vA}\right)^{-1}.\label{eq:tauh3}
\eeq
Note that (ignoring factors of $\sin\alpha$)
\beq
\frac{\tau_{H2}}{\tau_{H3}}\sim \chi.
\eeq
In this case with $\chi\gtrsim 1$, this process dominates until the RMHD-like terms on the RHS of Eq.~\ref{eq:els2ord} are comparable to the cubic nonlinear terms, 
which occurs at
\beq
\frac{z^+_2}{\vA}\sim \frac{z^-_2}{\vA}\sim \kperp d_i\frac{{z^+_1}^2}{\vA^2}.\label{eq:strongsat}
\eeq

\subsection{Coronal turbulence and laboratory experiments}
Since this nonlinear interaction is a factor of $\epsilon$ slower than the primary interaction between roughly-equal-amplitude counterpropagating waves, clearly it can only be important when $\zeta^-/\zeta^+\lesssim \epsilon$, i.e. the turbulence is rather imbalanced and the ``standard" RMHD nonlinearity is suppressed. While dispersive effects (e.g. at $\rho_s\nap \sim 1$) also allow for co-propagating interactions (see Sec. \ref{sec:1DVDAW} and Appendix \ref{app:rdaw}),  the mechanism in this section works even though linearizing the closed system comprised of Eqs. (\ref{eq:phi1}),(\ref{eq:psi1}) and (\ref{eq:els2ord}) results in non-dispersive AW, with $\omega=\pm \kpar \vA$. 

In the corona, $\beta\ll1$, justifying the use of Hall MHD, and moreover fluctuations are observed to be oblique and very imbalanced, the perfect situation for this interaction to become important. Moreover, the recently-discovered helicity barrier \citep{meyrand2021,squire2021} may cause the amplitude of the turbulence (and therefore $\epsilon$) to become rather large, helping the nonlinear effect here to become more important. 

Another situation where this calculation might be relevant is in laboratory experiments on LAPD, where $\kperp d_i \gg1$ and so the density fluctuations are relatively large. This decreases $\tau_{H2}$ (and $\tau_{H3}$) relative to $\tau_A$, and so this nonlinear interaction could be important when co-propagating waves with unaligned polarizations are launched in the device.

\section{One-dimensional steepening of dispersive Alfv\'en waves}\label{sec:1DVDAW}

The reason the co-propagating nonlinearity was a higher-order effect in Sec.~\ref{sec:3DRHMHD} was that the density fluctuations (compare  Eqs.~\ref{eq:densityintro} and \ref{eq:tauh2}) are still $O(\epsilon)$: in this section, we will study the case where we allow them to reach $O(1)$, by ordering $\kperp d_i \gg1$. Surprisingly we will find that large density fluctuations alone is not enough to allow AW steepening: even with $O(1)$ density fluctuations, a non-trivial one-dimensional nonlinear interaction of a single AW only occurs if one or more of the parameters $\kperp^2\rhot^2$, $\kperp^2 d_e^2$, $\kpar^2 d_i^2$, or $\eta \kperp^2/\omega$ are non-negligible: these are precisely those that cause the wave to be dispersive, see Eq.~\ref{eq:disp}.

\subsection{Basic equations}
To study nonlinear steepening, we may retreat to the comforting simplicity of one-dimensional dynamics. Unlike in Section \ref{sec:3DRHMHD}, we assume that all variation is in the $x$-direction, and that, without loss of generality, the background magnetic field lies in the $x-z$ plane, i.e. the full magnetic field vector is
\beq
\vb= (\vA\cos\theta,b_y,b_z+\vA\sin\theta),\label{eq:vacomp}
\eeq
which defines the angle $\theta$ and the fluctuating components of the magnetic field $b_y$ and $b_z$ (the $x$-component of the magnetic field is a constant by magnetic Gauss' law). Similarly, we write the ion velocity fluctuations in components $u_x, u_y, u_z$, and again write
\beq
n = 1 + \tn,
\eeq
so that $\tn$ is the normalized density fluctuation.
The two-fluid equations (\ref{eq:cty2}-\ref{eq:gauss2}) restricted to one dimension read
\begin{align}
    \ddt n &= -\ddx(nu_x),\label{eq:n1d}\\
    \Dt u_x &= -\frac{1}{n}\ddx\left(c_s^2 n+\frac{1}{2}b_y^2+\frac{1}{2}b_z^2+b_z\vA\sin\theta\right) - \frac{Zm_e}{m_i}\Dt u_x,\label{eq:ux1d}\\
    \Dt u_y &= \frac{1}{n}\vA\cos\theta \ddx b_y - \frac{Zm_e}{m_i}\Dt u_y - \frac{Zm_e}{m_i}d_i\Dt \frac{\ddx b_z}{n},\label{eq:uy1d}\\
    \Dt u_z &= \frac{1}{n}\vA\cos\theta \ddx b_z - \frac{Zm_e}{m_i}\Dt u_z + \frac{Zm_e}{m_i}d_i\Dt\frac{\ddx b_y}{n},\label{eq:uz1d}\\
    \Dt b_y &= \vA\cos\theta\ddx u_y - b_y\ddx u_x+d_i\ddx\Dt u_z+ \eta\ddx\left(\frac{\ddx b_y}{n}\right),\label{eq:by1d}\\
    \Dt b_z &= \vA\cos\theta\ddx u_z - b_z\ddx u_x - \vA\sin\theta\ddx u_x-d_i\ddx\Dt u_y\nonumber\\&\quad+ \eta\ddx\left(\frac{\ddx b_z}{n}\right)\label{eq:bz1d},
\end{align}
where $\Dt = \ddt + u_x\ddx$: because of the one-dimensional propagation, the only velocity driving nonlinear interaction is now the compressive velocity $u_x$. 

To derive a system that describes nonlinear dispersive Alfv\'enic fluctuations, we begin as in Sec.~\ref{sec:3DRHMHD}, assuming the waves are highly oblique and using as our expansion parameter
\beq
\epsilon \sim \cos\theta.\label{eq:epssec3}
\eeq
To keep the kinetic-Alfv\'en and inertial-Alfv\'en dispersive terms\footnote{Once again, keeping both of these simultaneously ignores strong Landau damping, and our results are really only valid when at least one of $d_e$ or $\rhot$ is small.} as well as resistivity, we will also order
\beq
\frac{c_s}{\vA}\sim \sqrt{\frac{Zm_e}{m_i}} \sim \nu_{ei} \sim \epsilon.
\eeq
To keep the finite-frequency effects, we order
\beq
\ddt \sim \vA\cos\theta\ddx \sim \Omega_i,
\eeq
which combined with (\ref{eq:epssec3}) gives
\beq
d_i\ddx \sim \epsilon^{-1},\label{eq:diddxbig}
\eeq
the ion inertial scale is large compared to the typical lengthscale of the fluctuations we are interested in. Combining the above equations, we have
\beq
d_i\cos\theta \ddx \sim \rhot\ddx \sim d_e\ddx \sim \eta\ddx^2\sim 1,\label{eq:orderingdisp}
\eeq
so that we are keeping all dispersive effects appearing in (\ref{eq:disp}). For the system to be strongly nonlinear, we require
\beq
\ddt \sim u_x\ddx (\sim \Omega_i),
\eeq
whence 
\beq
\frac{u_x}{\vA}\sim \epsilon.
\eeq
We will assume from the outset also that
\beq
\frac{u_y}{\vA}\sim \frac{b_y}{\vA}\sim \epsilon,\label{eq:order}
\eeq
showing that this is consistent later in the calculation.

The ordering of the density fluctuation $\tn$ follows from (\ref{eq:n1d}); ordering the different terms,
\beq
\tn\sim 1,
\eeq
and the density fluctuations can be large. The ordering for the final two variables, $b_z$ and $u_z$, may be deduced (similarly to how we obtained their size in Sec.~\ref{sec:3DRHMHD}) as follows. At lowest order in this expansion, (\ref{eq:ux1d}) becomes
\beq
\Dt u_x= -\frac{1}{n}\ddx\left(c_s^2 n+\frac{1}{2}b_y^2+b_z\vA\right),\label{eq:ux1d2}
\eeq
whence $b_z/\vA\sim \epsilon^2$ to be consistent with the other orderings.\footnote{We have preemptively used this to neglect the $b_z^2$ term; this also puts some weak restrictions on the ranges of $c_s/\vA$, $b_y/\vA$, and $\ddt/\Omega_i$ allowed, which we will not discuss.} Note that if $\Dt u_x$ and $(1/n)c_s^2 \nap\tn$ may be neglected, which are associated with dispersion at finite $\kpar d_i$ and $k\rhot$ respectively, all that is left is the MHD expression $b_z = -b_y^2/2\vA$ \citep{barnes1974,vasquez1996a}. Technically, this is already steepening, in that $b_z$ appears as a harmonic of $b_y$, but, when dispersive effects are relevant, the steepening we will derive shortly also occurs in $b_y$, and drives much larger-amplitude harmonics.

At lowest order, (\ref{eq:uz1d}) is
\beq
\Dt u_z = \frac{1}{n}\vA\cos\theta \ddx b_z + \frac{Zm_e}{m_i}d_i\Dt\frac{\ddx b_y}{n},\label{eq:uz1d2}
\eeq
and, to ensure consistency with the other orderings, $u_z/\vA\sim \epsilon^2$. 

These orderings mean that at lowest order, (\ref{eq:bz1d}) simply reads
\beq
\vA\ddx u_x = -d_i\ddx\Dt u_y.
\eeq
where we have used $\sin\theta=1-O(\epsilon^2)$.\footnote{This should be compared with the similar equation (\ref{eq:hpoldrift}) in the previous section.} Now, integrate once; the integration constant may be set to zero with either periodic or decaying boundary conditions, and we obtain
\beq
u_x = -\frac{1}{\Omega_i}\Dt u_y,\label{eq:poldrift}
\eeq
showing that our ordering of $u_x\sim u_y$ (\ref{eq:order}) is consistent. Note that (\ref{eq:poldrift}) shows that the compressive velocity results from the polarization drift \citep{hollweg1999}. In (\ref{eq:uy1d}), all we need to keep are
\beq
\Dt u_y = \frac{1}{n}\vA\cos\theta \ddx b_y,\label{eq:uy2}
\eeq
showing that $u_y\sim b_y$, as anticipated by (\ref{eq:order}). To make the equations more amenable to analysis, we first use (\ref{eq:poldrift}) in (\ref{eq:uy2}) to obtain
\beq
u_x = -\frac{1}{n}d_i\cos\theta\ddx b_y,\label{eq:ux3}
\eeq
which makes the way in which the dispersive parameter $d_i\cos\theta \ddx$ (i.e., $\kpar d_i$) controls the size of $u_x$ slightly more transparent. Using this in (\ref{eq:n1d}), we obtain an equation for the density
\beq
\ddt n = d_i\cos\theta\ddx^2 b_y.\label{eq:n3}
\eeq
Use (\ref{eq:ux1d2}) in (\ref{eq:uz1d2}) to eliminate $b_z$,
\begin{align}
\Dt u_z = &-\frac{1}{n}c_s^2\cos\theta\ddx n - \frac{1}{2n}\cos\theta\ddx b_y^2 \nonumber\\&- \cos\theta\Dt u_x + \frac{Zm_e}{m_i}d_i \Dt \frac{\ddx b_y}{n},\label{eq:uz3}
\end{align}
and then use (\ref{eq:uz3}) to eliminate $u_z$ from (\ref{eq:by1d}),
\begin{align}
\ddt b_y = &\vA\cos\theta\ddx u_y - \ddx(u_x b_y)\nonumber\\ &- d_i \cos\theta\ddx \left(\frac{1}{n}c_s^2\ddx n + \frac{1}{2n}\ddx b_y^2 + \Dt u_x\right),\nonumber\\&+d_e^2\ddx\Dt\frac{\ddx b_y}{n}+ \eta\ddx\frac{\ddx b_y}{n}.\label{eq:by3}
\end{align}
Using (\ref{eq:ux3}),
\beq
\frac{1}{2n}d_i\cos\theta\ddx b_y^2 = -u_x b_y,
\eeq
and using this in (\ref{eq:by3}) two of the nonlinear terms cancel, resulting in
\begin{align}
\ddt b_y = &\vA\cos\theta\ddx u_y - d_i \cos\theta\ddx \left(\frac{1}{n}c_s^2\ddx n + \Dt u_x\right)\nonumber\\&+d_e^2\ddx\Dt\frac{\ddx b_y}{n}+ \eta\ddx\frac{\ddx b_y}{n}.\label{eq:by4}
\end{align}
This last cancellation eliminates the usual terms that give rise to the DNLS (derivative nonlinear Schr\"odinger) type nonlinearity and associated solitons; the remaining nonlinearity is intrinsically dispersive, as we will show later. 

In summary, we have derived a closed set of model equations with which to study the physics of one-dimensional small-amplitude, strongly nonlinear, oblique, low-$\beta$ dispersive Alfv\'en waves; this set is
\begin{align}
u_x &= -\frac{1}{n}d_i\cos\theta\ddx b_y,\label{eq:ux1dmain}\\
\ddt n &= d_i\cos\theta\ddx^2b_y,\label{eq:dtn1dmain}\\
\Dt u_y &= \frac{1}{n}\vA\cos\theta\ddx b_y,\label{eq:dtuy1dmain}\\
\ddt b_y &= \vA\cos\theta\ddx u_y - d_i \cos\theta\ddx \left(\frac{1}{n}c_s^2\ddx n\right) \nonumber\\&\quad- d_i \cos\theta\ddx\Dt u_x+d_e^2\ddx\Dt\frac{\ddx b_y}{n}+ \eta\ddx\frac{\ddx b_y}{n}.\label{eq:dtby1dmain}
\end{align}
Equations equivalent to (\ref{eq:ux1dmain}-\ref{eq:dtby1dmain}) written in different variables were previously derived by Seyler \& Lysak (1999)\cite{seyler1999}. In the variables chosen here, the terms giving rise to dispersive waves (which we will call "dispersive terms" in this section) are all collected on the RHS of (\ref{eq:dtby1dmain}): from left to right, the terms on the RHS correspond to: (non-dispersive) Alfv\'enic propagation, dispersion due to finite $k \rhot$, $k_\parallel d_i$, $k d_e$, and resistive damping, respectively. However, each of these introduces not only linear dispersive terms but also additional nonlinearity. 

As an important caveat, this fluid model completely neglects many kinetic phenomena: for example, wave-particle interactions like Landau and cyclotron damping, and a realistic model of the ion response once $k \rho_i \sim 1$. It should therefore be used with caution and skepticism. For example, if $k d_e \sim k\rhot \sim 1$, the real system exhibits strong Landau damping\citep{zocco2011}, and even the real part of the linear frequency predicted by our equations is incorrect. It is therefore only reasonable to use these equations when at least one of $\kperp \rhot$ or $\kperp d_e$ is small. Likewise, the model ceases to be reasonable if $\kpar d_i$ becomes too large, in which case the real plasma exhibits strong ion cyclotron damping.
\subsection{Non-dispersive limit}\label{sec:nondisp}
First, we will briefly examine the equations in the strongly nonlinear but non-dispersive limit, i.e. setting the last four terms of (\ref{eq:dtby1dmain}) to zero. Let us try an Alfv\'en-wave solution, i.e. variables are functions of $x-t\vA\cos\theta$ (the opposite sign could be chosen if desired); then, $\ddt = -\vA\cos\theta\ddx$. Then, from (\ref{eq:dtn1dmain}),
\beq
\tn= -\frac{\ddx b_y}{\Omega_i}.\label{eq:nm1}
\eeq
(\ref{eq:dtby1dmain}) without the dispersive terms gives
\beq
b_y = - u_y. \label{eq:apos}
\eeq
Inserting (\ref{eq:ux1dmain}) into (\ref{eq:dtuy1dmain}), using $\ddt = -\vA\cos\theta\ddx$ and rearranging,
\beq
-\vA\cos\theta \ddx u_{y} = \frac{1}{n}\vA\cos\theta\left(\frac{\ddx b_y}{\Omega_i}\ddx u_y + \ddx b_y\right)
\eeq
Inserting (\ref{eq:nm1}) and (\ref{eq:apos}), we find that the nonlinear terms in this equation cancel, showing that the Alfv\'en wave is in fact a valid nonlinear solution no matter its (one-dimensional) spatial structure. Thus, if the dispersive terms are sufficiently small, to lowest order\footnote{If we were to formally continue this expansion to higher order in the dispersive parameters, we would find that the wave steepens at higher order.} the solution is a nonlinear Alfv\'en wave, with strongly nonlinear compressive flow and density fluctuations. The fact that the Alfv\'en wave survives so well in this regime agrees with the results already obtained in Sec.~\ref{sec:3DRHMHD} and in previous studies\citep{zocco2011}. This new calculation extends that result to larger $\kperp d_i$ (larger density fluctuations), with the important restriction to one dimension. In Mallet \emph{et al.} 2023\citep{mallet2023b}, we show that in fact this extension can continue all the way to $b_y/\vA\sim 1$.

The main subject of this calculation, of course, is to to study the nonlinear behaviour of these equations \emph{without} neglecting the dispersion. We will show that once the dispersion is significant, a monochromatic $b_y$ (for example) is no longer an exact solution, implying that the Alfv\'en wave is steepened due to the dispersive nonlinearity.
\subsection{Dispersive steepening}
Let us study the weakly nonlinear but strongly dispersive case, taking the overall amplitude of the fluctuations to be a factor $\delta\ll1$ smaller than in the original expansion in $\epsilon$ (Eq. \ref{eq:order})\footnote{These two expansions commute, but if any of the dispersive parameters becomes small in $\delta$ they must be dropped in the following analysis}, while keeping intact the ordering of $k d_i \sim \epsilon^{-1}$, as well as the order-unity dispersive parameters (Eq. \ref{eq:orderingdisp}). We then expand in powers of $\delta$:
\begin{align}
u_x &= u_{1x} + u_{2x} + \ldots,\nonumber\\
u_y&= u_{1y} + u_{2y} + \ldots,\nonumber\\
b_y &= b_{1y} + b_{2y} + \ldots,\nonumber\\
n &= 1+n_1 + n_2+\ldots,
\end{align}
where
\beq
\epsilon\vA \sim \frac{u_{1x}}{\delta} \sim \frac{u_{2x}}{\delta^2}, 
\eeq
and so on for the other variables. Qualitatively, one can already see the basic result we will obtain as follows. The ratio between each of the linear ($\delta$) and leading-order ($\delta^2$) nonlinear dispersive terms on the RHS of (\ref{eq:dtby1dmain}) is
\beq
n_1 \sim k d_i \frac{b_{1y}}{\vA}\sim \delta,\label{eq:n1scaling}
\eeq
which suggests that the harmonic amplitude will scale with $k d_i$, which is large. This will be confirmed by our mathematical analysis (see Eq.~\ref{eq:harmonic} later), and may explain why the harmonics observed on LAPD are quite substantial\citep{abler2023}; much larger than the MHD prediction \citep{barnes1974,vasquez1996a}.
\subsubsection{$O(\delta)$}
At first order, the equations are linear. Let us therefore assume sinusoidal variation, with fluctuations $\propto\exp(i(kx-\omega t))$. From (\ref{eq:ux1dmain}-\ref{eq:dtby1dmain}),
\begin{align}
u_{1x} &= -i\kpar d_i b_{1y},\quad
n_1 = - \frac{ik\kpar d_i}{\omega} b_{1y}, \quad
u_{1y} = -\frac{\kpar\vA}{\omega} b_{1y},\label{eq:linrels}\\
\omega b_{1y} &= -\kpar\vA u_{1y}+ik\kpar c_s^2d_i n_1 -i\omega \kpar d_i u_{1x} - k^2d_e^2\omega b_{1y} \nonumber\\&\quad - i\eta k^2 b_{1y},\label{eq:b1ylin}
\end{align}
where 
\beq
\kpar=k\cos\theta \sim \epsilon k\ll k.
\eeq
Substituting the linear relationships (\ref{eq:linrels}) into (\ref{eq:b1ylin}), we obtain the (expected) dispersion relation (\ref{eq:disp}).
\subsubsection{$O(\delta^2)$}
At second order, we must take into account the quadratic nonlinearities of the first-order quantities. It is obvious that the second-order fluctuations must therefore be proportional to $\exp(2i(kx-\omega t))$, and therefore do not in general lie on the linear DAW dispersion relation.\footnote{If more than a single sinusoid is considered at $O(\delta)$, one can also solve for the beat mode resulting in the difference of the phase functions. In our case with a single $O(\delta)$ wave, one just obtains $0=0$. We therefore do not consider this in the present analysis.}
From (\ref{eq:ux1dmain}-\ref{eq:dtuy1dmain}) we obtain
\begin{align}
    u_{2x} &= -2i\kpar d_i b_{2y} + i\kpar d_i n_1 b_{1y},\\
    n_2 &= -\frac{2ik\kpar d_i}{\omega}b_{2y},\\
    u_{2y} &= -\frac{\kpar\vA}{\omega}b_{2y} + \frac{1}{2}\frac{k}{\omega}u_{1x}u_{1y}+\frac{1}{2}\frac{\kpar\vA}{\omega}n_1b_{1y}.
\end{align}
Substituting the first-order relations (\ref{eq:linrels}) into the above equations, we write the right hand sides of the above equations in terms of $b_{2y}$ and $b_{1y}^2$ only,
\begin{align}
u_{2x} &= -2i\kpar d_i b_{2y}+ \frac{k\kpar^2d_i^2}{\omega}b_{1y}^2,\quad
n_2 = -\frac{2ik\kpar d_i}{\omega}b_{2y},\nonumber\\
u_{2y} &= - \frac{\kpar \vA}{\omega} b_{2y},\label{eq:nlrels}
\end{align}
where in the last equation the nonlinearity coming from the convective acceleration and the density inhomogeneities have cancelled. Equation (\ref{eq:dtby1dmain}) at second order reads
\begin{align}
-2i\omega b_{2y} = & 2i\kpar\vA u_{2y}
+4k\kpar d_i c_s^2 n_2 - 2k\kpar d_ic_s^2n_1^2 \nonumber\\
&-4 \omega\kpar d_i u_{2x} + 2 k\kpar d_i u_{1x}^2\nonumber\\
&+8i\omega k^2 d_e^2 b_{2y} - 4i\omega k^2d_e^2 n_1 b_{1y}-2ik^3d_e^2 u_{1x}b_{1y}\nonumber\\
&-4\eta k^2 b_{2y}+2\eta k^2 n_1 b_{1y},
\end{align}
We use the first-order (\ref{eq:linrels}) and second-order (\ref{eq:nlrels}) results to write the equation solely in terms of $b_{2y}$ and $b_{1y}^2$.
Solving for $b_{2y}$ and using (\ref{eq:disp}) to tidy up the expression, we obtain
\begin{widetext}
\beq
\frac{b_{2y}}{\vA} = ikd_i \frac{\kpar \vA}{\omega} \left(\frac{
k^2 \rhot^2 - 3 \kpar^2 d_i^2 -3 k^2 d_e^2 - \frac{i\eta k^2}{\omega} - 2k^2\rhot^2(\kpar^2d_i^2 + k^2 d_e^2)
}
{
3(\kpar^2 d_i^2 + k^2 d_e^2 - k^2 \rhot^2) + \frac{i\eta k^2}{\omega}-2k^2\rhot^2\frac{i\eta k^2}{\omega}
}
\right)\frac{b_{1y}^2}{\vA^2}
\label{eq:harmonic}
\eeq
\end{widetext}
This equation has some interesting properties. First, the driven harmonic amplitude $b_{2y}$ is of course proportional to $b_{1y}^2$, but it also scales with $k d_i$: this parameter is large in the ordering scheme (\ref{eq:diddxbig}), and so this could explain the somewhat significant harmonics observed in AW experiments on LAPD\citep{abler2023}. 

Second, the complicated fraction in between large brackets depends on the dispersion and resistive damping via the parameters $\kpar^2 d_i^2$, $k^2 d_e^2$, $k^2\rhot^2$, and $\eta k^2/\omega$. In (\ref{eq:ux1dmain}-\ref{eq:dtby1dmain}), all these parameters are considered to be order unity. However, (\ref{eq:harmonic}) shows that even if they are all somewhat small (provided that the $\delta$- and $\epsilon$- expansions both remain valid), the numerator and denominator will be dominated by the largest of these. Then, the dependence on top and bottom cancels out, giving a saturated harmonic amplitude that is independent of the value of the dispersive terms even in the weakly dispersive limit.


One serious limitation of this model is that it is one-dimensional, so that all wave modes must have the same propagation direction $\mathbf{k}/k$. This means that there are no RMHD-like interaction terms, for example, even if counterpropagating waves are put into the main equations (\ref{eq:dtn1dmain}-\ref{eq:dtby1dmain}), their wavevectors are perfectly aligned so the nonlinearity vanishes. It also means that the second harmonic is not in general a normal mode of the system (as we pointed out earlier).  

Finally, note that without the resistive damping, there would be a resonance at the point where the Alfv\'en wave dispersion vanishes (i.e. the positive and negative dispersive terms cancel). For this special value of the parameters, the coefficient of $b_{2y}$ vanishes: this may cause rather large harmonics for some values of the parameters. Probing this resonance would be a useful test of the model; however, as previously mentioned, care should be taken since if $\kperp d_e \sim \kperp \rhot \sim 1$ the dispersion relation (\ref{eq:disp}) is inaccurate, and the real system exhibits strong Landau damping.

The timescale for the nonlinear steepening to occur may be estimated in comparison with the non-dispersive timescale of the wave, i.e. we compare the timescale of  linear Alfv\'enic propagation (from the first term on the RHS of Eq.~\ref{eq:dtby1dmain}), $\tau_{A}^{-1}\sim \kpar\vA$ with the nonlinear dispersive terms (the four final terms on the RHS of Eq.~\ref{eq:dtby1dmain}). A different timescale may be estimated for each of the dispersive parameters $k^2\rhot^2, k^2 d_e^2, \kpar^2 d_i^2$, and the resistivity $\eta k^2/\omega$: the relevant nonlinear dispersive term in Eq.~\ref{eq:dtby1dmain} leads to the estimates
\beq
\tau_s^{-1}\sim n_1 \kpar\vA \times\max(k^2\rhot^2,
k^2 d_e^2,
\kpar^2 d_i^2,
\eta k^2/\omega),
\label{eq:taus}
\eeq
and so, using the estimate (\ref{eq:n1scaling}),
\beq
\frac{\tau_{A}}{\tau_s}\sim k d_i \frac{b_{y}}{\vA}\times\max(k^2\rhot^2,
k^2 d_e^2,
\kpar^2 d_i^2,
\eta k^2/\omega),
\eeq
where the largest dispersive parameter is chosen in the equations above and sets the steepening timescale. Thus, if the wave is strongly dispersive, i.e. the largest of these parameters is $O(1)$, the steepening is slower than the linear timescale $\tau_A$ by a factor $\delta$ in this weakly nonlinear case. However, if all the dispersive parameters are also small, the steepening is slower by a factor of $\delta$ times the largest dispersive parameter. Steepening can then only be significant if the system is allowed to evolve on this slow timescale; i.e. if other competing effects like the standard MHD-like counter-propagating nonlinearity or the three-dimensional interactions identified in  Section~\ref{sec:3DRHMHD} and Appendix~\ref{app:rdaw} are suppressed. We re-emphasize that,  according to Eq.~(\ref{eq:harmonic}), the saturated harmonic amplitude tends to a finite limit as the dispersive parameters become small, scaling only with $k d_i$: however, this will only be attained if the steepening is the fastest evolutionary process for the wave.

Our result shares some similarities with
Eq. (3.136) of Brugman 2007\cite{brugman2007}, who studied the nonlinear
interaction of two co-propagating DAW with aligned polarizations using
somewhat different ordering assumptions: their beat wave amplitude also scales with $k d_i$, and has a resonance at the point where the Alfven wave
dispersion vanishes.  Unlike our result, in which this resonance only
appears with cancellation of the positive and negative dispersive terms,
the Brugman expression incorrectly predicts this resonance will also
appear in the non-dispersive limit.  This is because they did not systematically keep all the relevant nonlinear terms: the cancellation of terms in Section~\ref{sec:nondisp} did not occur in this previous work, which would lead to an incorrect prediction of nonlinear steepening even in the non-dispersive limit.

\subsection{Coronal turbulence and laboratory experiments: comparison with Sec.~\ref{sec:3DRHMHD}}
As in Sec.~\ref{sec:3DRHMHD}, in the turbulent corona, this effect may become important because of the extreme imbalance. In laboratory experiments on LAPD, this effect could be important even when the dispersive corrections to the real frequency are small: the resistive damping can still be rather large for some choices of parameters\citep{bose2019}. Whether dispersive steepening (this section) or 3D nonlinear interaction (Sec.~\ref{sec:3DRHMHD}) dominates depends on how aligned or dispersive the waves are: for example, comparing (\ref{eq:taus}) with (\ref{eq:tauh2}),
\beq
\frac{\tau_s}{\tau_{H2}}\sim \frac{\sin^2\alpha}{\max(k^2\rhot^2,
k^2 d_e^2,
\kpar^2 d_i^2,
\eta k^2/\omega)},
\eeq
Thus, the scaling of turbulence dominated by these nonlinearities is likely to be highly non-universal, with the fastest timescale for co-propagating nonlinear interaction being one of $\tau_{H2}$, $\tau_{H3}$, or $\tau_s$ (with any of the dispersive terms), depending sensitively on parameters. In general, the scale at which these interactions will "switch on" scales with $\kperp d_i$, which may explain why a spectral break at this scale appears in the solar wind data at low $\beta$\citep{chen2014}, although further work is needed to develop a theory for turbulence dominated by these effects.
\subsection{Historical note: No small-amplitude AW solitons.}
If resistivity is negligible, one could attempt to find exact solutions to the original equations (\ref{eq:ux1dmain}--\ref{eq:dtby1dmain}) by looking for a solution with all variables travelling as a wave at speed $v$, i.e. $f=f(x-vt)$. Then, $\ddt = -v\ddx$. There are two potential classes of exact solutions; periodic waves (which at small amplitude behave as in the analysis above) and solitary waves, whose fluctuations tend to zero at infinity. Historically, there has been considerable debate as to the existence or non-existence of these small-amplitude AW solitons, with early analyses suggesting that kinetic and inertial AW solitons were possible at small amplitude\citep{hasegawa1976,shukla1982,wu1995}. The subject of Seyler \& Lysak (1999)\citep{seyler1999} (who, as mentioned previously, used without a detailed derivation our Eqs. \ref{eq:ux1dmain}--\ref{eq:dtby1dmain}) was in fact to show that all previous works neglected an important nonlinear term (the nonlinear polarization drift), and upon including this term, only singular solitary waves remain, necessarily containing a pair of density discontinuities. Our systematic derivation will hopefully convince an interested reader that the inclusion of this term is correct. In Mallet \emph{et al.} 2023\citep{mallet2023b}, we will treat the case of large-amplitude, slightly dispersive waves, and show that we recover both the small-amplitude singular solitons, but also new large-amplitude regular solitons.
\begin{table*}
\centering
\begin{tabular}{|c|c|c|c|}
\hline
     Section & Timescale symbol $\tau$ & $\tau_A/\tau$ & saturation amplitude $b_{2y}/\vA$ \\
\hline
Sec. \ref{sec:3wave}& $\tau_{H2}$&$(\kperp d_i b_{1y}/\vA)\sin^2\alpha$ [Eq.~\ref{eq:tauh2}] & $b_{1y}/\vA$\\
\hline
Sec. \ref{sec:strong}& $\tau_{H3}$&$(\kperp d_i b_{1y}/\vA)(\kperp b_{1y}/\kpar\vA)\sin\alpha$ [Eq.~\ref{eq:tauh3}]& $\kperp d_i b_{1y}^2/\vA^2$ [Eq.~\ref{eq:strongsat}]\\
\hline
Sec. \ref{sec:1DVDAW}& $\tau_{s}$&$\kperp d_i {b_{1y}}/{\vA}\times\max(\kperp^2\rhot^2,
\kperp^2 d_e^2,
\kpar^2 d_i^2,
\eta \kperp^2/\omega)$ [Eq.~\ref{eq:taus}]& $\kperp d_i b_{1y}^2/\vA^2$ [Eq.~\ref{eq:harmonic}]\\
\hline
\end{tabular}
\caption{The approximate nonlinear timescales and approximate saturated fluctuation amplitudes associated with different effects studied in the paper. Each nonlinear timescale is compared to the Alfv\'en time $\tau_A = (\kpar \vA)^{-1}$, while the saturation amplitude normalized to the Alfv\'en velocity $b_{2y}/\vA$ is expressed in terms of the amplitude of the primary wave(s) $b_{1y}$. The interaction in Section \ref{sec:3wave} is resonant, and only saturates once it reaches an amplitude comparable to the primary wave, at which point the ordering assumptions break down. The saturation amplitude in Sec. \ref{sec:1DVDAW} also technically depends on the values of the dispersive parameters (see Eq.~\ref{eq:harmonic}); roughly speaking, this amounts to a prefactor of order unity correcting the entry in this table. $\alpha$ is a typical angle between perpendicular wavevectors of interacting waves (see Sec.~\ref{sec:3DRHMHD}).}\label{tab:summarydisc}
\end{table*}

\section{Discussion}\label{sec:disc}
In this paper, we have analysed the nonlinear properties of small-scale two-fluid Alfv\'en waves. Linearly, highly oblique two-fluid AW have two main differences from MHD. First, when $\kperp d_i\sim 1$, they develop a significant density fluctuation\citep{hollweg1999} due to the Hall effect,
\beq
\frac{\delta n_i}{n_{0i}} \sim - i \kperp d_i \frac{\delta B}{B_0},\label{eq:densitydisc}
\eeq
Second, they become dispersive, with the dispersion controlled by the parameters $\kperp^2 \rhot^2$, $\kperp^2 d_e^2$ and $\kpar^2 d_i^2$ (see Eq.~\ref{eq:disp}). Our nonlinear analysis uncovers two new and interesting phenomena that stem from these new effects.

In Sec.~\ref{sec:3DRHMHD}, we study a regime where $\kperp d_i\sim 1$ in three dimensions, but we consider non-dispersive AW (i.e, we take $\rhot \to 0$ and $d_e \to 0$, and $\omega/\Omega_i \sim \epsilon$), leaving the equations of compressible Hall MHD. Using a small-amplitude, anisotropic, low-frequency expansion of these equations, we show that at the lowest non-trivial order we recover reduced MHD, for which co-propagating AW with arbitrary spatial structure are an exact, non-dispersive nonlinear solution. However, at the second non-trivial order we show that there is a resonant nonlinear interaction between co-propagating waves with non-aligned wavevectors: even though the linear waves still obey the MHD linear dispersion relation $\omega=\pm \kpar \vA$. Importantly, this nonlinearity vanishes if the perpendicular wavevectors in the system are aligned, and so a single sinusoidal wave does not self-interact via this process. We estimate the timescale for this process (\ref{eq:tauh2} and \ref{eq:tauh3}). From these timescales, it is clear that the important parameter controlling the co-propagating interaction timescale is the size of the AW's density fluctuations (Eq.~\ref{eq:densitydisc}). Numerical simulations of the decay of 2D large-amplitude Alfv\'en wavepackets have recently been performed by Tenerani \emph{et al.} 2023 \cite{tenerani2023}, with observed timescale for the interaction $\tau^*\sim \tau_A (l/d_i)$: these simulations were performed for fixed $\delta B/B_0 \sim l_\perp/l_\parallel \sim 1$. Inserting these fixed factors into the expressions for our timescales, one finds that, indeed, $\tau^*\sim\tau_{H2}\sim\tau_{H3}$, so the interaction derived analytically in this paper may extend to large-amplitude, less oblique waves.

In Section~\ref{sec:1DVDAW}, we use a different ordering scheme, allowing these density fluctuations to be large, $\delta n / n_0 \sim 1$, and also retain the terms that give rise to dispersive waves at lowest order (Eq.\ref{eq:disp}) in a two-fluid model. We show that, even with such large density fluctuations, due to a fortuitous cancellation in the equations nonlinear steepening in one dimension is absent if the terms giving rise to dispersion are neglected, only becoming significant once at least one of the terms is non-negligible, as evidenced by an expression for the timescale of this nonlinear steepening process (Eq.~\ref{eq:taus}), which is inversely proportional to the dispersive parameters. We also derive an explicit formula (\ref{eq:harmonic}) for the amplitude of the harmonic of a primary wave that would be produced by this mechanism; unlike the timescale, this is only mildly dependent on the dispersion: weakly dispersive waves take longer to evolve but have comparable saturated harmonic amplitudes. Both the harmonic amplitude and the inverse of the timescale are additionally proportional to $\kperp d_i$ since each nonlinearity is driven by the density fluctuations (Eq.~\ref{eq:densitydisc}). 
A related model which can deal with very-large-amplitude waves with $\delta B/B_0\sim 1$ and solitons is developed in Mallet \emph{et al.} 2023\citep{mallet2023b}.

We summarize the main physical implications of the new nonlinear interactions identified in in Sections \ref{sec:3DRHMHD} and \ref{sec:1DVDAW}, in terms of their fully-saturated amplitudes and dynamical timescales, in Table \ref{tab:summarydisc}. The saturated amplitudes can be quite large: in Section \ref{sec:3DRHMHD} the interaction is resonant, and continues until the ordering assumptions break down, and in Section \ref{sec:1DVDAW} the harmonic amplitude scales with $\kperp d_i$, which can be large. Conversely, the timescales associated with both interactions are typically slow compared to the timescale associated with the interaction of two roughly equal-amplitude counter-propagating AW, even at $\kperp d_i\sim 1$. However, there are two important applications for which the new nonlinearities are likely to be highly relevant.

First, the strength of the nonlinear interactions identified here scales with $\kperp d_i$ (equivalently, with the density fluctuations). This means that care must be taken when interpreting the results of laboratory AW experiments: typically on LAPD, $\kperp d_i\gg1$. Our results may provide a way to understand the generation of wave harmonics on LAPD, as well as understanding interactions between multiple co-propagating AW.

Second, the fluctuations in the solar corona and inner solar wind, currently being explored by Parker Solar Probe and Solar Orbiter, are dominated by large-amplitude, outward-travelling Alfv\'en waves, with only a relatively small inward-travelling component. At large scales, this means that the usual counter-propagating nonlinear interaction is suppressed, and the nonlinear physics that we have studied here may be important, especially at sharp switchback boundaries\citep{krasnoselskikh2020}. Moreover, a recent exciting theoretical result shows that the dominant nonlinear cascade is further suppressed due to a "helicity barrier"\citep{meyrand2021,squire2021}, which causes the turbulence amplitude to increase until the usual small-amplitude, anisotropic approximation breaks down, allowing the effects here to become even more significant. Further work will study the complex interplay of the three-dimensional interaction of Sec.~\ref{sec:3DRHMHD} and the one-dimensional dispersive steepening of Secs.~\ref{sec:1DVDAW}, and their potential application to turbulence in the corona. One important point is that, unlike in the MHD counter-propagating interaction, these nonlinearities couple waves to larger $\kpar$, meaning that they may help the turbulence access ion cyclotron heating entering at $\kpar d_i\sim 1$, thought to be important for the ion heating necessary to accelerate the solar wind\citep{hollweg2002}.

\acknowledgements
AM is grateful to C. Chaston, J. Bonnell, and S. Boldyrev for useful discussions. A.M. was supported by NASA grant 80NSSC21K0462 and NASA contract NNN06AA01C. S.D. was supported by DOE grant DE-SC0021237
and NASA grant 80NSSC18K1235. M.A. was supported by DOE grants DE-SC0021291 and DE-SC0023326. T.B. was supported by NASA grant 80NSSC21K1771. C.H.K.C. was supported by UKRI Future Leaders Fellowship MR/W007657/1 and STFC Consolidated Grant ST/T00018X/1.
\appendix

\section{Two-fluid model}\label{app:2f}
Here, we discuss the basic set of model two-fluid equations that are the basis for the calculations in the main body of the paper. We begin with isothermal two-fluid equations for electrons and ions,
\begin{align}
\ddt n_i + \nabla\cdot(n_i \vu) &= 0,\label{eq:cty1}\\
Zm_e n_i(\ddt \vu_e +\vu_e\cdot\nabla\vu_e) &= -Zen_i\left(\vE+\frac{\vu_e\times\vB}{c}\right) \nonumber\\&\phantom{=}\, - ZT_e\nabla n_i - \nu_{ei} Z n_i m_e(\vu_e-\vu),\label{eq:emom}\\
m_in_i(\ddt \vu + \vu\cdot\nabla\vu) &= {Zen_i}\left(E+\frac{\vu\times\vB}{c}\right)\nonumber\\&\phantom{=} - {T_i}\nabla n_i - \nu_{ie}n_i m_i(\vu-\vu_e),\label{eq:imom}\\
\nabla\times\vB &= \frac{4\pi}{c}\vJ,\label{eq:ampere}\\
\vJ &= Zen_i(\vu-\vu_e),\label{eq:Jdef}\\
\ddt \vB &= -c\nabla\times\vE,\label{eq:faraday}\\
\nabla\cdot\vB &= 0,\label{eq:gauss}
\end{align}
where notations are standard except that we have denoted the ion velocity as simply $\vu$, and we are assuming quasineutrality, $n_e=Zn_i$. We first use (\ref{eq:emom}) to obtain
\begin{align}
\vE = &- \frac{\vu_e\times\vB}{c} - \frac{m_e}{e}\left(\ddt \vu_e + \vu_e\cdot \nabla\vu_e\right)\nonumber\\ &-\frac{T_e}{e}\nabla\ln (n_i/n_{0i})+ \frac{m_e}{Ze^2 n_i}\nu_{ei}\vJ.
\end{align}
In this equation, we then use (\ref{eq:Jdef}) to eliminate $\vu_e$, and then Amp\`ere's law (\ref{eq:ampere}) to eliminate $\vJ$ in favour of $\vB$, obtaining
\begin{align}
\vE = &-\frac{\vu\times\vB}{c} + \frac{1}{4\pi Z e}\frac{(\nabla\times\vB)\times\vB}{n_i}-\frac{m_e}{e}\Dt\vu\nonumber\\&+\frac{cm_e}{4\pi Ze^2}\Dt\left(\frac{\nabla\times\vB}{n_i}\right)+\frac{cm_e}{4\pi Ze^2}\nu_{ei}\frac{\nabla\times\vB}{n_i}
\nonumber\\&-\frac{T_e}{e}\nabla\ln (n_i/n_{0i}) \nonumber\\&+ \frac{cm_e}{4\pi Ze^2}\frac{\nabla\times\vB}{n_i}\cdot\nabla\left(\vu - \frac{c}{4\pi Z e}\frac{\nabla\times\vB}{n_i}\right),\label{eq:ohms}
\end{align}
where we have denoted $\Dt=\ddt+\vu\cdot\nabla$. We insert (\ref{eq:ohms}) into the ion momentum equation (\ref{eq:imom}), noting that the collisional drag term vanishes by symmetry, and thus obtain
\begin{align}
\Dt\vu &= \frac{1}{4\pi m_i}\frac{(\nabla\times\vB)\times\vB}{n_i} - \frac{Zm_e}{m_i}\Dt\vu \nonumber\\
&+ \frac{cm_e}{4\pi e m_i}\Dt\left(\frac{\nabla\times\vB}{n_i}\right) - c_s^2\nabla\ln (n_i/n_{0i})\nonumber\\
&+\frac{cm_e}{4\pi e m_i}\frac{\nabla\times\vB}{n_i}\cdot\nabla\left(\vu - \frac{c}{4\pi Z e}\frac{\nabla\times\vB}{n_i}\right),\label{eq:imom2}
\end{align}
where 
\beq
c_s^2 = \frac{T_i+ZT_e}{m_i}.
\eeq
Another expression for $\vE$ may be found from (\ref{eq:imom}), again using (\ref{eq:Jdef}) and (\ref{eq:ampere}) to eliminate $\vu_e$,
\beq
\vE = \frac{m_i}{Z e}\Dt u - \frac{\vu\times\vB}{c}+\frac{T_i}{Ze}\nabla\ln (n_i/n_{0i}) + \frac{\nu_{ie} m_i c}{4\pi Ze^2 }\frac{\nabla\times\vB}{n_i}.\label{eq:ohms2}
\eeq
Using (\ref{eq:ohms2}) in (\ref{eq:faraday}),
\beq
\ddt \vB = \nabla\times(\vu\times\vB) - \frac{m_i c}{Ze}\nabla\times\Dt\vu - \frac{\nu_{ie} m_i c^2}{4\pi Z^2 e^2}\nabla\times\left(\frac{\nabla\times\vB}{n_i}\right).\label{eq:ind1}
\eeq
Equations (\ref{eq:cty1}),(\ref{eq:imom2}), (\ref{eq:ind1}) along with (\ref{eq:gauss}) are the complete set required to determine the evolution; to make them more transparent, we normalize the density and magnetic field according to
\beq
n=\frac{n_i}{n_{0i}}, \quad \vb = \frac{\vB}{\sqrt{4\pi n_{0i}m_i}}.
\eeq
Additionally employing the definitions of the ion and electron inertial scales
\beq
d_i = \frac{c\sqrt{m_i}}{\sqrt{4\pi Z^2 e^2 n_{0i}}}, \quad d_e = \frac{c\sqrt{m_e}}{\sqrt{4\pi e^2 n_{0e}}}= \sqrt\frac{Zm_e}{m_i} d_i,
\eeq
and the Ohmic resistivity
\beq
\eta = \nu_{ei}d_e^2 = \nu_{ie}d_i^2,
\eeq
the equations may be written
\begin{align}
    \ddt n &=- \nabla\cdot(n\vu),\label{eq:cty2}\\
    \Dt \vu &= \frac{1}{n}\vb\cdot\nabla\vb - \frac{1}{n}\nabla \left(c_s^2 n + \frac{1}{2}b^2\right)+ \frac{Z m_e}{m_i} d_i \Dt \frac{\nabla\times\vb}{n} \nonumber\\&\quad- \frac{Zm_e}{m_i}\Dt\vu + \frac{Zm_e}{m_i}d_i\frac{\nabla\times\vb}{n}\cdot\nabla\left(\vu - d_i\frac{\nabla\times\vb}{n}\right),\label{eq:mom2}\\
    \Dt \vb &= \vb\cdot\nabla\vu - \vb \nabla\cdot \vu - d_i\nabla\times\Dt\vu - \eta\nabla\times\left(\frac{\nabla\times\vb}{n}\right),\label{eq:ind2}\\
    \nabla\cdot\vb &= 0.\label{eq:gauss2}
\end{align}
These equations, in different limits, are analysed in the main sections of this paper. This fluid model obviously neglects many important plasma processes, for example, wave-particle interaction resulting in Landau/cyclotron damping. This means that our model is not always appropriate: for example, when $\kpar d_i \sim 1$ or when $\kperp \rho_i \sim 1$: future work will incorporate these effects.

We focus on the Alfv\'enic modes, and take
\beq
\ddt \sim \vA \ddz \lesssim \Omega_i,
\eeq
which orders out the fast mode, and $c_s/\vA=\sqrt{\beta} \ll 1$, which removes the slow mode. We also assume oblique waves, $\ddz /\nap \ll 1$. With these assumptions, one obtains the dispersion relation\citep{hollweg1999}
\beq
(1 + k_\parallel^2d_i^2 + k^2 d_e^2)\omega^2 + i\eta k^2\omega - k_\parallel^2\vA^2(1+k^2\rhot^2) = 0,\label{eq:dispapp}
\eeq
with solution
\beq
\omega = -i\gamma_0 \pm \sqrt{\omega_0^2-\gamma_0^2},
\eeq
where
\beq
\omega_0 = k_\parallel \vA \sqrt{\frac{1+k^2\rhot^2}{1+k_\parallel^2d_i^2+k^2d_e^2}},\quad \gamma_0 = \frac{1}{2}\frac{\eta k^2}{1+k_\parallel^2 d_i^2 + k^2 d_e^2}.
\eeq
As it results from a fluid model, this dispersion relation is a poor approximation to reality in a number of important situations. For example, when both $ k \rhot \sim 1$ and $k d_e \sim 1$, real AW undergo strong Landau damping, completely ignored in this model; even the real frequency in this regime also differs significantly from this two-fluid model\citep{thuecks2009}. For $k \rho_s \sim 1$ but $k d_e \ll1$ and $k \rho_i \ll 1$ (fluid kinetic AW) and $k d_e \sim 1$ but $k \rhot \ll 1$ (fluid inertial AW), the collisionless damping is relatively weak, and (\ref{eq:dispapp}) is somewhat acceptable. 

\section{Reduced dispersive AW equations}\label{app:rdaw}

Evidence suggests that the small-scale turbulence in the solar wind\citep{chen2013} consists of kinetic AW; in the aurora, both inertial and kinetic AW exist in different regions \citep{stasiewicz2000}. Modelling of this turbulent cascade is often done assuming that typical linear and nonlinear frequencies are small compared to $\Omega_i$\citep{seyler2003,schektome2009,zocco2011}. To make contact with these existing theories, we use the following ordering scheme. First, we write the magnetic field $b$ as
\beq
\vb = \vA + \vbperp + \hat{\vz}b_z,
\eeq
where $\vA$ is the mean Alfv\'en velocity and $\vbperp$ and $b_z$ are the (small-amplitude) magnetic fluctuations in velocity units in the perpendicular and $z$ directions. Similarly, we write
\beq
n = 1+ \tn,
\eeq
where $\tn$ is the normalized density fluctuation. 

We assume that the perpendicular lengthscales of our fluctuations are small compared to their parallel lengthscales, and expand in $\epsilon\ll1$, defined via
\beq
\frac{\ddz}{\nap} \sim \epsilon^2\ll1.
\eeq
We also assume that the linear timescales are Alfv\'enic, and comparable to the nonlinear timescales and the electron-ion collision frequency, i.e.
\beq
\ddt \sim \vA\ddz \sim \vuperp\cdot \nap \sim \vbperp \cdot \nap \sim \nu_{ei},
\eeq
which implies that the fluctuation amplitudes are ordered as
\beq
\frac{u_\perp}{\vA}\sim \frac{\bperp}{\vA} \sim \epsilon^2 \ll1.
\eeq
We also restrict ourselves to low $\beta$ and small mass ratio, taking
\beq
\beta = \frac{c_s^2}{\vA^2}\sim \frac{Z m_e}{m_i}\epsilon^2.
\eeq
We wish to include the kinetic and inertial Alfv\'en dispersion, so we must order\footnote{Again (cf. discussion in Appendix \ref{app:2f}), keeping both terms simultaneously is a poor approximation, but we will here do it anyway to compare with previous work.}
\beq
\rhot \nap \sim d_e\nap \sim 1,\label{eq:keepdisp}
\eeq
implying that the resistive term is also kept: since $\eta = \nu_{ei}d_e^2$, \beq
\eta \nap^2 \sim \ddt.
\eeq
Eq.~(\ref{eq:keepdisp}) also means that $d_i$ is large compared to the perpendicular scales,
\beq
d_i\nap \sim \epsilon^{-1},
\eeq
but small compared to the parallel scales,
\beq
d_i\ddz \sim \epsilon,
\eeq
a regime typical for auroral fluctuations and in some laboratory experiments. This also means that the typical timescales are small compared to $\Omega_i$,
\beq
\frac{1}{\Omega_i}\ddt \sim \epsilon.
\eeq
We will assume from the outset (and show that this is consistent) that $b_z$ and $u_z$ are at least as small as the perpendicular fluctuations $\vuperp$ and $\vbperp$; we will deduce the ordering for $\tn$ from the equations, but assume it is small compared to unity. The $z$-component of the induction equation (\ref{eq:ind2}) at lowest order yields
\beq
\nap\cdot\vuperp = - \frac{1}{\Omega_i}\nap \times \Dt \vuperp \sim \epsilon^2\Omega_i,\label{eq:divcurl}
\eeq
meaning that the perpendicular velocity is only slightly compressive: its divergence is one order smaller than its curl. This can be inserted into the continuity equation (\ref{eq:cty2}), which then gives at lowest order
\beq
\Dt \tn = \frac{1}{\Omega_i}\nap \times \Dt \vuperp,\label{eq:ncurl}
\eeq
from which
\beq
\tn \sim \epsilon.
\eeq
Turning now to the perpendicular momentum equation (\ref{eq:mom2}), at lowest order we just obtain the pressure balance
\beq
c_s^2 \tn = - b_z\vA,\label{eq:pbalance}
\eeq
whence in fact, we see that
\beq
\frac{b_z}{\vA} \sim \epsilon^3.
\eeq
The lowest order terms of the $z$-component of the momentum equation (\ref{eq:mom2}) are
\begin{align}
\Dt u_z = &\vA\ddz b_z + \vbperp\cdot\nap b_z + \frac{Zm_e}{m_i}d_i \Dt \nap\times\vbperp \nonumber\\&+ d_e^2 [\hat{\vz}\times\nap b_z] \cdot \nap \nap \times\vbperp,\label{eq:uz}
\end{align}
from which we can deduce that\footnote{We have used this to preemptively drop the term involving $u_z$ on the RHS of (\ref{eq:mom2}) when writing (\ref{eq:uz}).}
\beq
\frac{u_z}{\vA} \sim \epsilon^3.
\eeq
At lowest order, the perpendicular induction equation reads
\beq
\Dt \vbperp = \vA\ddz\vuperp +\vbperp\cdot\nap\vuperp + d_i \hvz\times\nap\Dt u_z + \eta \nap^2\vbperp,
\eeq
into which we insert (\ref{eq:uz}) to eliminate $u_z$ and (\ref{eq:pbalance}) to eliminate $b_z$ in favor of $\tn$, yielding
\begin{align}
\ddt \vbperp = &\vA\ddz\vuperp + \nap\times(\vuperp\times\vbperp) \nonumber\\&- \frac{c_s^2}{\Omega_i}\hvz\times\nap\left[\vA\ddz \tn + \vbperp\cdot\nap \tn\right] \nonumber\\
&- d_e^2\frac{c_s^2}{\Omega_i}\hvz\times\nap[\hvz\times\nap \tn \cdot \nap \nap\times\vbperp]\nonumber\\
&+d_e^2\hvz\times\nap \Dt\nap\times\vbperp+\eta\nap^2\vbperp.\label{eq:perpind}
\end{align}
To make further progress, note that since (\ref{eq:gauss2}) at lowest order is just $\nap \cdot \vbperp = 0 + O(\epsilon^3)$, we may write $\vbperp$ in terms of a flux function
\beq
\vbperp = \hvz \times \nap \psi.\label{eq:psidef}
\eeq
Similarly, because $\nap \cdot \vuperp$ is small (cf. \ref{eq:divcurl}), at lowest order we can write the perpendicular velocity in terms of a stream function\footnote{We have not completely neglected the compressive part of the velocity; it sets the density fluctuations via (\ref{eq:divcurl}): in Secs. \ref{sec:3DRHMHD} and \ref{sec:1DVDAW}, it plays a more active role in the dynamics and we must include it explicitly.}
\beq
\vuperp = \hvz\times\nap\phi.
\eeq
Nonlinear terms are expressed in terms of these scalar functions as
\beq
\vuperp \cdot \nap f = \brak{\phi}{f}, \quad \vbperp \cdot \nap f =\brak{\psi}{f}
\eeq
where the Poisson bracket $\brak{f}{g}= \nap f \times \nap g$. To lowest order,
\beq
\Dt f= \ddt f+ \brak{\phi,f}.
\eeq
Writing (\ref{eq:ncurl}) in terms of $\phi$ results in
\beq
\tn = \frac{1}{\Omega_i}\nap^2\phi,\label{eq:nphi}
\eeq
the density fluctuations are proportional to the vorticity of the flow. Writing (\ref{eq:perpind}) and the curl of the perpendicular momentum equation (\ref{eq:mom2}) in terms of $\phi$ and $\psi$ only, we obtain a closed set of nonlinear equations for the DAW,
\begin{align}
\ddt \nap^2\phi + \{\phi,\nap^2\phi\} = &\vA\ddz\nap^2\psi + \{\psi,\nap^2\psi\},\label{eq:vort}\\
\ddt[\psi - d_e^2\nap^2\psi] = &\vA\ddz [\phi -\rhot^2\nap^2\phi] \nonumber\\&+\{\psi - d_e^2\nap^2\psi,\phi-\rhot^2\nap^2\phi\}\nonumber\\&+\eta \nap^2\psi.\label{eq:psi}
\end{align}
These "reduced dispersive Alfv\'en wave" (RDAW) equations are identical to those used in Seyler \& Xu (2003)\citep{seyler2003} to study auroral fluctuations; however, those authors did not provide a systematic derivation\footnote{To the author's knowledge, one does not appear in press.}. The equations are also similar to the reduced gyrokinetic equations in Zocco \& Schekochihin 2011\citep{zocco2011} (a.k.a. "KREHM"): however, that model is significantly more sophisticated, incorporating a rigorous treatment of the ion and electron kinetics.

Here, we will just make two important points about the above equations. First, the Poisson brackets all vanish if gradients are all in one direction: so any one-dimensional configuration of fields is an exact solution; i.e., there is no one-dimensional wave steepening retained in this system. Second, if both $\rhot \nap \ll1$ and $d_e\nap\ll 1$, the RVDAW equations become reduced magnetohydrodynamics (RMHD), for which $\phi=\pm \psi$ are exact nonlinear solutions, travelling in opposite directions along $\hvz$, regardless of three-dimensional structure. In this non-dispersive limit, non-trivial nonlinear interaction only occurs between counterpropagating waves. This is despite the density fluctuations (\ref{eq:nphi}) and despite $d_i\nap \gg 1$: to this order, the compressive nature of the wave makes no difference to the nonlinear dynamics. Upon including the dispersive effects, nonlinear interaction does occur between co-propagating waves (provided their perpendicular wavevectors also point in different directions) with different perpendicular scales, because smaller-scale waves "catch up" to larger-scale ones.

\bibliography{mainbib2}

\end{document}